\documentclass[12pt]{iopart}

\usepackage{setspace}
\usepackage{graphicx}
\usepackage{epstopdf}
\usepackage{enumerate}
\usepackage{indentfirst}
\usepackage[figbotcap]{subfigure}
\usepackage{slashed}
\usepackage{braket}
\usepackage{setstack}
\usepackage{mathrsfs,amssymb}

\usepackage[left=1.54cm,top=1.54cm,bottom=1.54cm,right=1.54cm]{geometry}
\renewcommand{\theequation}{\arabic{equation}}  
\date{\today}
\usepackage[usenames,dvipsnames]{color}

\newcommand \beq{\begin{eqnarray}}
\newcommand \eeq{\end{eqnarray}}

\begin{document}

\title[Hyperfine structure of the hydroxyl free radical (OH) in electric and magnetic fields]
{Hyperfine structure of the hydroxyl free radical (OH) in electric and magnetic fields}

\author{Kenji Maeda$^1$, Michael L Wall$^2$, and Lincoln D Carr$^1$}

\address{$^1$Department of Physics, Colorado School of Mines, Golden, Colorado 80401, USA }
\address{$^2$JILA, NIST and Department of Physics, University of Colorado, Boulder, CO 80309, USA }
\ead{kenji.bosefermi@gmail.com}
\begin{abstract}
We investigate single-particle energy spectra of the hydroxyl free radical (OH) 
in the lowest electronic and rovibrational level under combined static electric and magnetic fields, 
as an example of heteronuclear polar diatomic molecules. 
In addition to the fine-structure interactions, 
the hyperfine interactions and centrifugal distortion effects are taken into account 
to yield the zero-field spectrum of the lowest ${}^2\Pi_{3/2}$ manifold to an accuracy of less than 2\,kHz. 
We also examine level crossings and repulsions in the hyperfine structure 
induced by applied electric and magnetic fields. 
Compared to previous work, 
we found more than 10 percent reduction of the magnetic fields 
at level repulsions in the Zeeman spectrum subjected to a perpendicular electric field. 
It is important to take into account  hyperfine structure 
when we investigate physics of OH molecules at micro-Kelvin temperatures and below.


\end{abstract}

\maketitle

\section{Introduction}
\label{sec:introduction}

The hydroxyl free radical (OH molecule) is 
a simple but fascinating molecule in various fields of science;  
chemistry, astronomy, and physics. 
In chemistry, the OH molecule was the first short-lived molecule to be investigated by 
microwave spectroscopy and by gas-phase electron paramagnetic resonance (EPR) 
\cite{Dousmanis_1955,Radford_1961}, 
and was also the first free radical to be detected in a molecular beam \cite{Meulen_1972}. 
In astronomy, it was found that the OH molecule was the first interstellar molecule detected at radio frequencies 
\cite{Weinreb_1963,Weaver_1965}. 
The OH/IR stars, which exhibit OH line emission bright at near infrared wavelength, 
yielded so much intensity as OH sources that they led to an interpretation based on maser amplification, 
inspiring the concept of astrophysical masers \cite{Elitzur_1982}. 
In the field of cold and ultracold physics, the hydroxyl free radical has received a renewed attention 
as a constituent of quantum dipolar systems. 
A gas of OH molecules has been recently 
trapped and evaporatively cooled to milli-Kelvin temperatures at JILA \cite{Ye_2012_2}. 

Although there are many studies of the single-particle spectrum of the OH molecule 
in the context of chemistry and radio-astronomy as mentioned above, 
in the presence of electric and magnetic fields the energy spectra of OH 
have been calculated previously only to energy scales of milliKelvin temperatures (or frequencies of megaHertz) 
\cite{Ye_2012_2,Brown_1978,Brown_1981,Coxon_1980,
Bohn_2002,Bohn_2005,Ye_2012_1,Bohn_2013_1,Bohn_2013_2,Bhattacharya_2013}, 
far from the sub-microKelvin temperatures at which OH molecules will become quantum degenerate. 
Therefore, as a necessary step towards a correct description of 
a quantum degenerate molecular dipolar gas of OH molecules at sub-microKelvin temperatures, 
we investigate its hyperfine structure in the lowest
electronic and rovibrational states under combined electric and magnetic fields. 
In addition to the fine-structure interactions, 
we fully consider the hyperfine interactions and centrifugal distortion effects 
to obtain the zero-field spectrum of the lowest ${}^2\Pi_{3/2}$ manifold 
to an accuracy of less than $2\,{\rm kHz}\sim100\,{\rm nK}$. 
We also examine level crossings and repulsions in hyperfine structure 
in the presence of applied electric and magnetic fields to 
explore how these level crossings and repulsions change 
when we change the relative angle between the electric and magnetic fields.

Ahead of ultracold molecules, 
ultracold gases of atoms with magnetic dipole moments, or atomic dipolar gases, 
have been intensely investigated. 
Gases of Chromium \cite{Pfau_2008}, 
Dysprosium \cite{Lev_2011}, 
and Erbium \cite{Aikawa_2012} 
have been trapped and cooled down to quantum degeneracy in experiments. 
Due to their anisotropic long-range dipole-dipole interactions, 
atomic dipolar gases are expected to exhibit novel quantum phenomena: 
spin textures \cite{Yi_2006,Kawaguchi_2007,Vengalattore_2008}, dipolar relaxation \cite{Pasquiou_2010}, 
Einstein-de Haas effects \cite{Kawaguchi_2006,Santos_2006} in their bosonic spieces, 
and ferronematic \cite{Miyakawa_2008,Fregoso_2009_1} and  
antiferrosmectic-C phases \cite{Maeda_2013} in their fermionic species. 
However, dipole-dipole interactions between atoms are fixed in strength 
by their permanent magnetic dipole moments. 
In contrast, polar molecules 
offer an electric dipole moment that is directly tunable via an applied electric field and can be made orders of magnitude larger 
than dipole moments in atoms. 
Cold and ultracold gases of molecules with electric dipole moments moments -- molecular dipolar gases -- 
are at are at or rapidly approaching quantum degeneracy in experiments, 
e.g., KRb \cite{Ye_2013}, RbSr \cite{Pasquiou_2013}, OH \cite{Ye_2012_2}, and SrF \cite{Shuman_2010}. 
Recently, it has been shown that 
the single-particle spectrum and the dipole-dipole interactions of magnetic dipoles 
in a magnetic field can be simulated by symmetric top molecules subject to an electric field \cite{Wall_2013_1}, 
and we stress that 
the hydroxyl radical (OH molecule), the subject of this article, 
also has the symmetric top structure in its electric dipole moment.

This paper is organized as follows. 
Section~2 is a preliminary section, overviewing a theoretical background of the OH free radical. 
In section~3, the hyperfine structure of OH is explored in the absence of external fields. 
We first introduce the effective Hamiltonian for the OH molecule and 
numerically diagonalize it to obtain the energy spectrum of OH in the lowest energy manifold. 
Section~4 deals with the effects of applied electric and magnetic fields, 
that is, the Stark and Zeeman effects in OH. 
We calculate the Stark and Zeeman spectra independently, and then 
examine the energy spectra in the presence of both electric and magnetic fields at relative angles of
 $0^{\circ}$, $45^{\circ}$, and $90^{\circ}$. 
We compare our results to previous work 
and show that the coupling with the excited states gives non-negligible contributions 
to the values of magnetic fields where level repulsions occur. 
Section~5 is devoted to the summary and concluding remarks. 
Appendix~A summarizes definitions and formulae for the spherical tensor operators. 
Appendix~B gives matrix elements of the effective Hamiltonian for OH 
in the Hund's case (a) basis.

\section{Overview of the hydroxyl radical}
\begin{figure}[t]
\begin{center}
\includegraphics[width=12cm]{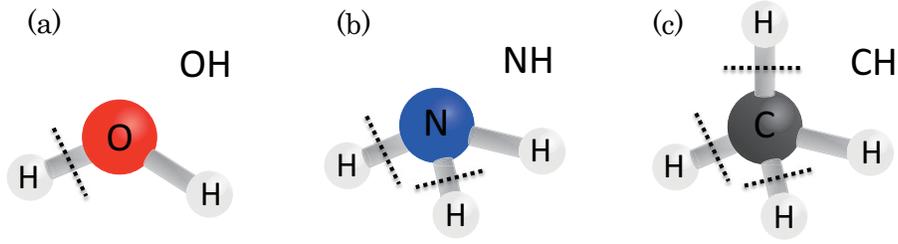}
\end{center}
\vspace{-0.5cm}
\caption{Free radicals: (a) OH, (b) NH, and (c) CH. 
The dotted lines represent dissociation of one or more hydrogen atoms to make free radicals from 
chemically stable, non-reactive, molecules. 
}
\label{FIG_free_radicals}
\end{figure}

We begin by introducing the hydroxyl (OH) free radical, 
for those readers who may be unfamiliar with it or find useful a reminder of its theoretical framework at higher temperatures 
($\gtrsim\!\!10^3\,{\rm K}$). 
The OH molecule can be obtained by dissociation of a hydrogen atom from a water molecule ${\rm H_2O}$, 
as shown in Figure~\ref{FIG_free_radicals}(a). 
Thus, the OH molecule has a chemically reactive, or radical, electron in its open shell. 
Other free radicals, like the NH and CH radicals shown in Figures~\ref{FIG_free_radicals}(b) and (c), 
are also easily obtained from stable molecules, ${\rm NH}_{3}$ (ammonia) and ${\rm CH}_{4}$ (methane), respectively. 
Our model of the OH molecule is composed of an oxygen nucleus, a hydrogen nucleus, and nine electrons. 
The total kinetic energy operator is given by 
\beq
\hat{T}&=&
-\frac{\nabla_{_{\rm O}}^2}{2M_{_{\rm O}}}
-\frac{\nabla_{_{\rm H}}^2}{2M_{_{\rm H}}}
-\sum_{i=1}^{9}\frac{\nabla_{i}^2}{2m_{\rm e}}
~. 
\eeq
Here $M_{_{\rm O}}$, $M_{_{\rm H}}$, and $m_{\rm e}$ 
are masses of the oxygen nucleus, hydrogen nucleus, and electron, respectively, 
and $\nabla_{_{\rm O}}$, $\nabla_{_{\rm H}}$, and $\nabla_{i}$ are gradient operators 
with respect to the coordinates of the oxygen nucleus ${\bf r}_{_{\rm O}}$, 
hydrogen nucleus ${\bf r}_{_{\rm H}}$, and electrons ${\bf r}_i$, respectively. 
Throughout this paper, we shall choose units such that $\hbar=1$ unless quoting an energy in frequency units. 
Setting the center of mass of the oxygen nucleus and hydrogen nucleus as the origin of the coordinate system, 
we rewrite the kinetic energy operator as
\beq
\hat{T}&=&
-\frac{\nabla_{_{\rm M}}^2}{2M}
-\frac{\nabla_{\rm r}^2}{2m}
-\sum_{i=1}^{9}\frac{\nabla_{i}'^2}{2m_{\rm e}}
-\frac{1}{2(M_{_{\rm O}}+M_{_{\rm H}})}
\sum_{i,j=1}^{9}  \nabla_{i}'\cdot\nabla_{j}'
~,    
\eeq
with the total mass of the molecule $M=M_{_{\rm O}}+M_{_{\rm H}}+9m_{\rm e}$ 
and the reduced mass of the two nuclei $m=M_{_{\rm O}}M_{_{\rm H}}/(M_{_{\rm O}}+M_{_{\rm H}})$. 
Note that $\nabla_{_{\rm M}}$, $\nabla_{\rm r}$, and $\nabla_{i}'$ 
are gradient operators with respect to the center of mass coordinate of the molecule 
${\bf r}_{_{\rm M}}=(M_{_{\rm O}}{\bf r}_{_{\rm O}}+M_{_{\rm H}}{\bf r}_{_{\rm H}}+\sum_{i=1}^{9}m_{\rm e}{\bf r}_i)/M$, 
the relative coordinate of two nuclei ${\bf r}={\bf r}_{_{\rm H}}-{\bf r}_{_{\rm O}}$, 
and electron coordinates ${\bf r}_i'$ measured from the center of mass of the two nuclei. 
For the relative motion of the oxygen nucleus and hydrogen nucleus ${\bf r}$, 
we use the polar coordinate representation $(r,\theta,\phi)$, which gives
\beq
\hat{T}&=&
-\frac{\nabla_{_{\rm M}}^2}{2M}
-\frac{1}{2mr^2}\frac{\partial}{\partial r}\biggl(r^2\frac{\partial}{\partial r}\biggl)
+\frac{1}{2mr^2}\hat{\bf R}^2
-\sum_{i=1}^{9}\frac{\nabla_{i}'^2}{2m_{\rm e}}
-\frac{1}{2(M_{_{\rm O}}+M_{_{\rm H}})}
\sum_{i,j=1}^{9}  \nabla_{i}'\cdot\nabla_{j}'
~, 
\label{mol_kin}
\eeq
with the square of the rotational angular momentum operator of two nuclei given by 
\beq
\hat{\bf R}^2
&=&
-\frac{1}{\sin\theta}\frac{\partial}{\partial \theta}\biggl(\sin\theta\frac{\partial}{\partial \theta}\biggl)
-\frac{1}{\sin^2\theta}\frac{\partial^2}{\partial \phi^2}~.
\eeq
The first term in Eq.~(\ref{mol_kin}) represents the translational motion of the molecule, 
that will be neglected in our single molecule study, 
the second and third terms describe the radial and rotational motion of two nuclei 
related to vibration and rotation of the molecule, 
and the forth and fifth terms are the kinetic energy operators of the electrons 
and the mass polarization term, respectively. 
Due to the large mass difference between electrons and nuclei, 
$m_{\rm e} \ll M_{_{\rm O}}, M_{_{\rm H}}$, we will neglect the mass polarization term 
and define the electronic Hamiltonian of the OH molecule with the Coulombic potential as
\beq
\hat{H}_{\rm el}&=&
-\sum_{i=1}^{9}\frac{\nabla_{i}'^2}{2m_{\rm e}}
+\frac{e^2}{4\pi\epsilon_0}\left\{
\sum_{i<j}\frac{1}{r_{ij}}-\sum_{i=1}^{9}\left(\frac{Z_{_{\rm O}}}{r_{_{\rm O}i}}+\frac{Z_{_{\rm H}}}{r_{_{\rm H}i}}\right)
+\frac{e^2}{4\pi\epsilon_0}\frac{Z_{_{\rm O}}Z_{_{\rm H}}}{r}
\right\}
~,
\eeq
where $r_{ij}$ is the distance between electrons $i$ and $j$, and
$r_{_{\rm O}i}$ ($r_{_{\rm H}i}$) is the distance between the electron $i$ 
and the oxygen nucleus (hydrogen nucleus). 
The electric charges of electrons, the oxygen nucleus and hydrogen nucleus 
are $e$, $Z_{_{\rm O}}e$, and $Z_{_{\rm H}}e$, respectively, 
and $\epsilon_0$ is the permittivity of vacuum. 
Since we included the Coulombic potential between the oxygen nucleus and hydrogen nucleus 
in the electronic Hamiltonian, the nuclear Hamiltonian of the OH molecule is just given 
by the kinetic energy operator of the relative motion of the nuclei 
\beq
\hat{H}_{\rm nucl}&=&
-\frac{1}{2mr^2}\frac{\partial}{\partial r}\biggl(r^2\frac{\partial}{\partial r}\biggl)
+\frac{1}{2mr^2}\hat{\bf R}^2
~. 
\eeq

Using a singular perturbation method in terms of a mass-ratio parameter $(m_{\rm e}/m)^{1/4}$, 
Born and Oppenheimer \cite{BO_1927} showed that in a typical molecule 
$\Delta E_{\rm vib}/\Delta E_{\rm el}\approx  \Delta E_{\rm rot}/\Delta E_{\rm vib} \approx(m_{\rm e}/m)^{1/2}$, 
where $\Delta E_{\rm el}$, $\Delta E_{\rm vib}$, and $\Delta E_{\rm rot}$ 
are the energetic separations between the ground state and the first excited state 
in electronic, vibrational, and rotational energy levels, respectively\footnote{
A simple estimate is as follows \cite{Baym_1969}. 
From the uncertainty principle, 
in a molecule of size $a$ electrons have momenta typically of $\hbar/a$, and we have 
$\Delta E_{\rm el}\sim \hbar^2/(2m_{\rm e} a^2)$. 
The vibrational motion of nuclei distorts the electron wave functions, 
and its approximate harmonic oscillator potential $m\omega^2 a^2/2$ 
should be on the order of the electronic excitation energy, 
which gives the typical frequency of the vibration $\omega\sim (m_{\rm e}/m)^{1/2}\hbar/(m_{\rm e} a^2)$. 
The nuclear vibration is thus spaced in energy by $\hbar\omega$, which gives the relation 
$\Delta E_{\rm vib}/\Delta E_{\rm el}\sim (m_{\rm e}/m)^{1/2}$. 
Finally, the rotational energy of nuclei with angular momentum $\hbar N$ is estimated as 
$\hbar^2N(N+1)/(2I)$ with the momentum of inertia of nuclei $I\sim ma^2$, 
yielding $\Delta E_{\rm rot}/\Delta E_{\rm vib}\sim (m_{\rm e}/m)^{1/2}$. 
}. 
This general result separates electronic and nuclear motions in a molecule 
and allows us to solve the Schr\"{o}dinger equation for electrons 
in the presence of the electrostatic field produced by fixed nuclear charges, 
\beq
\hat{H}_{\rm el}\,\psi_{\rm el}^{(n)}&=&
E_{\rm el}^{(n)}\psi_{\rm el}^{(n)}. 
\label{BO1}
\eeq
Here we note that since the internuclear distance $r$ is fixed in Eq.~(\ref{BO1}) 
the resultant eigenvalues $\{E_{\rm el}^{(n)}(r)\}_n$ and eigenfunctions $\{\psi_{\rm el}^{(n)}({\bf r}'_1,\dots , {\bf r}'_9;r)\}_n$ 
depend on $r$ parametrically. 
For the Schr\"{o}dinger equation of the whole OH molecule, 
\beq
(\hat{H}_{\rm el}+\hat{H}_{\rm nucl})\Psi&=&E\Psi
\label{Schro_OH}
~, 
\eeq
we assume a partial separation of variables of the molecular wave function in the form of 
\beq
\Psi({\bf r}'_1,\dots , {\bf r}'_9,r,\theta ,\phi)&=&
\sum_n 
\psi_{el}^{(n)}({\bf r}'_1,\dots , {\bf r}'_9;r) \chi^{(n)} (r,\theta ,\phi)
~.
\eeq
We defined the nuclear part of the molecular wave function by $\chi^{(n)}$. 
Then the molecular Schr\"{o}dinger equation, Eq.~(\ref{Schro_OH}), reduces to
\beq
\left(
\frac{\hat{p}_r^2}{2m}+\frac{1}{2mr^2}\hat{\bf R}^2+E_{\rm el}^{(n)}-E
\right) \chi^{(n)}
&=&\sum_{m}\hat{K}_{mn}  \chi^{(m)} 
~, 
\label{BO2}
\eeq
with matrix elements $\hat{K}_{mn}$ which couple different electronic states, 
\beq
\hat{K}_{mn}&=&
\int{\rm d}{\bf r}_1\cdots {\rm d}{\bf r}_9\, 
\psi^{(m)\ast}_{\rm el}
\left\{
\frac{1}{m}\left(\hat{p}_r\psi^{(n)}_{\rm el}\right)\hat{p}_r+\frac{1}{2m}\left(\hat{p}_r^2\psi^{(n)}_{\rm el}\right)
\right\}~,
\eeq
where $\hat{p}_r$ is the momentum operator conjugate to $r$, 
\beq
\hat{p}_r&=&
\frac{1}{ir}\frac{\partial}{\partial r} r
~. 
\eeq
In the {\it Born-Oppenheimer approximation}, 
which corresponds to the lowest order approximation in $(m_{\rm e}/m)^{1/2}$ \cite{BO_1927,Baym_1969}, 
all the matrix elements $\hat{K}_{mn}$ are set to be zero, 
and $\chi^{(n)}$ becomes an eigenfunction of
\beq
\left(
\frac{\hat{p}_r^2}{2m}+\frac{1}{2mr^2}\hat{\bf R}^2+E_{\rm el}^{(n)}
\right) \chi^{(n)}
&=& E \chi^{(n)} 
~, 
\label{BO3}
\eeq
where $E_{\rm el}^{(n)}$ can be considered as an {\it adiabatic potential} for the vibration of nuclei 
in a given electronic state.

Even if we fix the internuclear distance, 
the electronic Schr\"{o}dinger equation (\ref{BO1}) cannot be solved exactly. 
Therefore, we usually model the exact solutions of Eq.~(\ref{BO1}) 
using the electron wave functions in an isolated oxygen atom 
and those in an isolated hydrogen atom. 
This ansatz is called the {\it linear combination of atomic orbitals (LCAO) method} 
since the electron wave functions in the molecule are modeled by 
linear combinations of the electron wave functions in the parent atoms, called {\it atomic orbitals}. 
We shall start from a review on atomic orbitals for the sake of completeness.  
The wave functions of an electron in the hydrogen atom are given by 
$\psi^{\rm H}_{n\ell m_{\ell}}({\bf r}_1)=R_{n\ell}(r_1)Y_{\ell m_{\ell}}(\theta_1,\phi_1)$ 
with the associated Laguerre functions $R_{n\ell}$ and spherical harmonics $Y_{\ell m_{\ell}}$. 
The quantities $n$, $\ell$, and $m_{\ell}$ are the principal, 
azimuthal, and magnetic quantum numbers, respectively, having the allowed vales: 
$n\in\mathbb{N}$, $\ell=0,1,2, \dots , n-1$, and $m_{\ell}=0,\pm1,\pm2, \dots , \pm l$. 
The electron wave functions $\psi^{\rm H}_{n\ell m_{\ell}}$ are known as atomic orbitals, 
and for $\ell=0,1,2,3,\dots$, they are referred as $s$, $p$, $d$, $f$ orbitals, respectively. 
The values of the principal quantum numbers $n$ are specified as prefixes, 
$1s$, $2s$, $2p$, $3s$, $3p$, $3d$, and so forth. 
As for the magnetic quantum numbers $m_{\ell}$, 
we take real linear combinations of the complex wave functions $\psi^{\rm H}_{n\ell m_{\ell}}$ 
with different vales of $m_{\ell}$. For example, 
we define $\psi^{\rm H}_{2p_{x}}=(\psi^{\rm H}_{2,1,1}+\psi^{\rm H}_{2,1,-1})/\sqrt{2}$ and  
$\psi^{\rm H}_{2p_{y}}=-i(\psi^{\rm H}_{2,1,1}-\psi^{\rm H}_{2,1,-1})/\sqrt{2}$, 
and call them $2p_x$ and $2p_y$ orbitals 
because they are real functions and have their maxima along $x$- and $y$-directions, respectively. 
The $2p_z$ orbital is given by $\psi^{\rm H}_{2p_{z}}=\psi^{\rm H}_{2,1,0}$ 
since the wave function $\psi^{\rm H}_{2,1,0}$ is a real function with its maximum along the $z$-axis. 
On the other hand, the electron wave functions, or atomic orbitals, of the oxygen atom cannot be given exactly 
since we cannot analytically solve the Schr\"{o}dinger equation for eight electrons interacting with each other 
via the Coulombic potential. 
Then we usually consider the {\it Hartree-Fock variational state} $\psi^{\rm O}_{\rm HF}$ 
\cite{Hartree_1928a,Hartree_1928b,Fock_1930a,Fock_1930b} 
composed of a simple product of one-electron wave functions $\phi_i$, also called {\it orbitals}, 
and fully antisymmetrized with respect to interchange of any two electrons, 
\beq
\!\!\!\!\!\!\!
\psi^{\rm O}_{\rm HF}({\bf r}_1,s_1,{\bf r}_2,s_2,\dots , {\bf r}_8,s_8)
&=&
\frac{1}{\sqrt{8!}}\sum_{\sigma\in\mathfrak{S}_8}{\rm sgn} (\sigma) \phi_{\sigma(1)}({\bf r}_1,s_1)
 \phi_{\sigma(2)}({\bf r}_2,s_2)\cdots \phi_{\sigma(8)}({\bf r}_8,s_8)
 ~,~~
\eeq
where $\mathfrak{S}_8$ stands for the symmetric group of eight elements
and ${\rm sgn} (\sigma)$ is the sign of the permutation $\sigma$ taking its value 
$1$ and $-1$ for even and odd permutations, respectively. 
Note here that we specify components of the spin-1/2 spinors of electrons by $s_i=\pm1/2$, 
assuming that one-electron wave functions can be decomposed into 
$\phi_i({\bf r}_i,s_i)=\varphi_i({\bf r}_i)\chi_i(s_i)$ with spatial orbitals $\varphi_i$ and spin orbitals $\chi_i$. 
Minimizing the energy expectation in $\psi^{\rm O}_{\rm HF}$ subject to the condition that 
$\{\phi_i\}_{i=1}^{8}$ is an orthonormal set
gives the coupled nonlinear equations for the oxygen atomic orbitals, 
\beq
&&
\left(-\frac{\nabla_{i}^2}{2m_{\rm e}}-\frac{Z_{\rm O}e^2}{4\pi\epsilon_0}\frac{1}{|{\bf r}_i|}
+\sum_{j\neq i}\sum_{s_j=\pm1/2}\int{\rm d}{\bf r}_j\,\frac{e^2}{4\pi\epsilon_0}\frac{|\phi_j({\bf r}_j,s_j)|^2}{|{\bf r}_i-{\bf r}_j|}
\right)\phi_i({\bf r}_i,s_i)
\nonumber\\
&&
-\sum_{j\neq i}\sum_{s_j=\pm1/2}\int{\rm d}{\bf r}_j\,\frac{e^2}{4\pi\epsilon_0}\frac{1}{|{\bf r}_i-{\bf r}_j|}
\phi^{\ast}_j({\bf r}_j,s_j)\phi_j({\bf r}_i,s_i)\phi_i({\bf r}_j,s_j)
~=~
\varepsilon_i \phi_i({\bf r}_i,s_i)
~, 
\label{HF}
\eeq
which are called the {\it Hartree-Fock equations}. 
The Hartree-Fock equations (\ref{HF}) consist of, in order, 
the kinetic energy of an electron, the Coulombic potential between the electrons and the oxygen nucleus, 
the averaged repulsions from the other electrons, 
the so-called {\it exchange potential} arising from the antisymmetrization in $\psi^{\rm O}_{\rm HF}$, 
or from the Pauli exclusion principle in electrons, 
and the energy of the orbital $\phi_i$. 
Numerical solutions of the Hartree-Fock equations (\ref{HF}), usually obtained by iteration, 
supports the {\it shell model of atoms}. 
The shell model has the following major features: 
(i) the quantum numbers of atomic orbitals are given by $n$, $\ell$, $m_{\ell}$ like in the hydrogen atom, 
and also by spin quantum number $m_s$, i.e., $\phi_i({\bf r}_i,s_i)=\varphi_{n\ell m_{\ell}}({\bf r}_i)\chi_{m_s}(s_i)$; and 
(ii) the energy of the orbital $\varepsilon_i$ depends on $n$ and $\ell$ due to the deviation from a pure $1/r$-potential, 
but not on $m_{\ell}$ and $m_s$, thus we have $4\ell +2$ degenerate orbitals for given values of $n$ and $\ell$. 
This set of $4\ell +2$ orbitals is called a {\it shell}, and the distribution of electrons 
in these shells is called the {\it electron configuration}. 

\begin{figure}[t]
\begin{center}
\includegraphics[width=12cm]{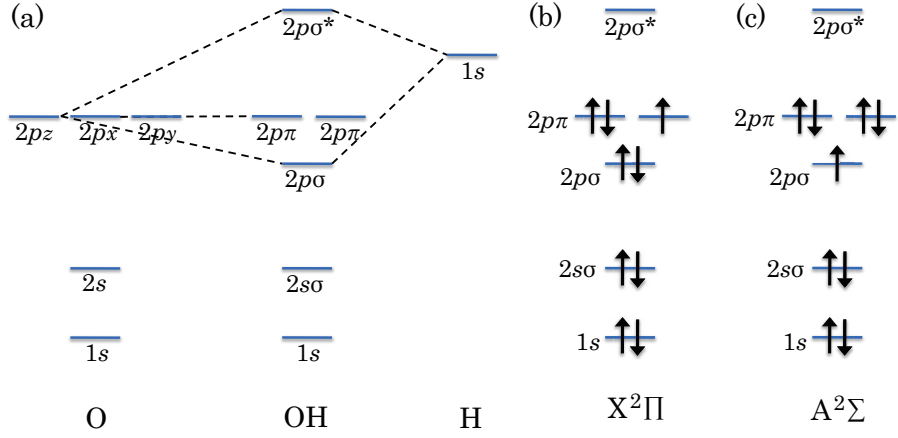}
\end{center}
\vspace{-0.5cm}
\caption{(a) Energy level diagram of electrons in OH molecule based on the LCAO method. 
(b) Electronic configuration in the $X{}^{2}\Pi$ ground state, 
and (c) in the first excited $A{}^{2}\Sigma$ state of OH molecule, 
where the up and down arrows occupying the molecular orbitals 
denote electrons with spin quantum numbers $1/2$ and $-1/2$, respectively. 
}
\label{ES_OH}
\end{figure}

Next we discuss the qualitative nature of the electron wave functions of molecules, called {\it molecular orbitals}. 
According to the LCAO method, 
the OH molecule has the molecular orbitals and their enegy levels for electrons, as shown in Figure~\ref{ES_OH}(a). 
The $1s$ and $2s$ atomic orbitals of the oxygen atom are energetically far separated from 
the $1s$ atomic orbital of the hydrogen atom, 
and only the $2p$ atomic orbitals of the oxygen atom can couple with the $1s$ atomic orbital of the hydrogen atom. 
Among the $2p$ atomic orbitals of the oxygen atom, 
the $2p_z$ orbital has overlap with the $1s$ atomic orbital of the hydrogen atom, 
while the $2p_x$ and $2p_y$ orbitals do not have overlap in spatial average 
and they are renamed as $2p\pi$ molecular orbitals. 
Then, the $2p_z$ orbital of the oxygen atom and $1s$ orbital of the hydrogen atom 
interact in-phase to form a bonding molecular orbital $2p\sigma$, 
or out-of-phase to form an antibonding molecular orbital $2p\sigma^{\ast}$. 
Note that in the $\sigma$ and $\pi$ molecular orbitals, 
an electron has the projection of orbital angular momentum along the internuclear axis $0$ and $\pm1$, respectively. 
Thus, the lowest energy electron configuration of the OH molecule, called the $X{}^{2}\Pi$ state,
 is given by 
\beq
X{}^{2}\Pi&:& (1s)^2(2s\sigma)^2(2p\sigma)^2(2p\pi)^3~,
\eeq
as shown in Figure~\ref{ES_OH}(b), and the first excited electronic state, called the $A{}^{2}\Sigma$ state, 
becomes 
\beq
A{}^{2}\Sigma&:& (1s)^2(2s\sigma)^2(2p\sigma)^1(2p\pi)^4~,
\eeq
as shown in Figure~\ref{ES_OH}(c). 
Here the superscripts of the braces represent the numbers of electrons occupied 
taking into account electronic spin degrees of freedom. 
Defining $\Lambda$ as  the projection of electronic orbital angular momentum along the 
symmetry axis of  the molecule, 
we have $\Lambda=\pm1$ in the $X{}^{2}\Pi$ electronic state, 
while $\Lambda=0$ in the $A{}^{2}\Sigma$ electronic state. 
The two-fold degeneracies in states with $|\Lambda|\neq 0$ are called {\it $\mathit{\Lambda}$-doubling}, 
which comes from the fact that the Hamiltonian of a diatomic molecule is invariant under a reflection 
with respect to a plane containing the symmetry axis of the molecule 
while $\Lambda$ changes its sign under such reflection. 
The electronic ground $X{}^{2}\Pi$ state and the first excited $A{}^{2}\Sigma$ state 
are far separated from each other by $4\,{\rm eV}\sim10^3\,{\rm THz}\sim 5\times 10^{4}\, {\rm K}$, 
which is the largest energy scale in the hierarchy of energy levels. 
Fuller details on electronic structures of molecules can be found in \cite{Slater_1963}. 

Now the electronic structures in the molecule 
become input parameters of the nuclear Schr\"{o}dinger equation (\ref{BO3}). 
Since in the Born-Oppenheimer approximation there is no mixing of different electronic states, 
we can write down the ground state of the OH molecule as
$\Psi=\psi_{el}^{(X{}^{2}\Pi)}\chi^{(X{}^{2}\Pi)}$, and 
the nuclear wave function $\chi^{(X{}^{2}\Pi)}$ is determined by Eq.~(\ref{BO3}) 
with the $X{}^{2}\Pi$ adiabatic potential $E^{(X{}^{2}\Pi)}$. 
Noting that there is the hierarchy in energy, $ \Delta E_{\rm rot}/\Delta E_{\rm vib} \approx(m_{\rm e}/m)^{1/2}$,
we can first investigate the vibrational motion assuming that the rotational motion of nuclei is frozen out. 
For a stable molecule the electron configuration of the ground state supports 
the chemical bond of the molecule, and therefore the adiabatic potential $E^{(X{}^{2}\Pi)}$ should have its minimum 
at the {\it equilibrium distance} $r_{0}$ of the nuclei. 
Employing the {\it harmonic approximation}, that is, replacing $E^{(X{}^{2}\Pi)}$ 
by the first two terms in a power series expansion about $r=r_0$, 
we extract the vibrational motion from Eq.~(\ref{BO3}), 
\beq
-\frac{1}{2mr}\frac{{\rm d}^2}{{\rm d} r^2}\{rf_v(r)\}
+\frac12k(r-r_0)^2 f_v(r)
&=&
E_{v}f_v(r)
~,
\label{BO4}
\eeq
where $k$ is a curvature given by the harmonic approximation $E^{(X{}^{2}\Pi)}(r)\simeq k(r-r_0)^2/2$. 
We can analytically solve Eq.~(\ref{BO4}) to obtain the eigenfunctions 
$ f_v(r)=c_v r H_v(\tilde{r}) e^{-\tilde{r}^2/2}$ and eigenenergies $E_v=(v+1/2)\omega_{\rm vib}$, 
where $H_v$ is the Hermite polynomial with vibrational quantum number $v=0,1,2,\dots$, and 
the harmonic vibrational frequency is given by $\omega_{\rm vib}=\sqrt{k/m}$. 
We also defined a dimensionless variable $\tilde{r}=\sqrt{m\omega_{\rm vib}}(r-r_0)$ 
and a normalization constant $c_v$. 

For the rest of the motion in Eq.~(\ref{BO3}), the rotational motion of nuclei, 
we assume that the internuclear distance is fixed to the equilibrium or average value $r_0$. 
Then, the rotational motion of nuclei reduces to motion of a {\it rigid rotor} 
\beq
\frac{\hat{\bf R}^2}{2mr_0^2}Y_{R,M_R}(\theta,\phi)&=& E_{R,M_R}Y_{R,M_R}(\theta,\phi)~,
\eeq
which immediately gives the spherical harmonics $Y_{R,M_R}$ as eigenfunctions 
and their eigenenergies as $E_{R,M_R}=R(R+1)/(2mr_0^2)$ 
where $R=0,1,2,\dots$ and $M_R=0,\pm1,\pm2\dots,\pm R$. 
Note that due to centrifugal forces of rotational motion 
the internuclear distance will increase from $r_0$ as the molecule rotates faster. 
This modifies the moment of inertia in Eq.~(\ref{BO4}) and 
gives corrections to the rigid rotor energies, 
called {\it centrifugal distortion corrections}\footnote{
We can get an insight into centrifugal distortion corrections based on classical mechanics \cite{BrownCarrington_2003}. 
Suppose that in the rotating molecule with its angular velocity $\omega$, 
the internuclear distance increases to $r_{_{\rm CD}}$ 
which is determined by the requirement that the centrifugal force $m\omega^2 r_{_{\rm CD}}$ 
is balanced by the restoring force $k(r_{_{\rm CD}}-r_{0})$, 
that is, $r_{_{\rm CD}}=[1+m\omega^2/(k-m\omega^2)]r_0$. 
The rotational energies are modified to 
$E=R_{cl}^2/(2mr_{_{\rm CD}}^2)+k(r_{_{\rm CD}}-r_{0})^2/2$ 
where $R_{cl}=m r_{_{\rm CD}}^2 \omega$ is the classical angular momentum. 
Assuming the stiffness of the molecule ($k\gg m\omega^2$), 
we obtain $E\simeq R_{cl}^2/(2mr_{0}^2)-R_{cl}^4/(2mr_{0}^6k)$ 
to the lowest order in $m\omega^2/k$, 
and the term proportional to $R_{cl}^4$ is a centrifugal distortion correction. 
}. 
It is in part the centrifugal distortion corrections which are vital to include in order to reach $100\,{\rm nK}$ precision for the OH molecule. 
Treating nuclear motion under the harmonic oscillator and rigid rotor approximations, 
we can write down the nuclear wave functions in the electronic ground state as 
$\chi_{v,R,M_R}^{(X{}^{2}\Pi)}(r,\theta,\phi)=f_v(r)Y_{R,M_R}(\theta,\phi)$ 
with their eigenenergies $E_{v,R,M_R}=(v+1/2)\omega_{\rm vib}+R(R+1)/(2mr_0^2)$.

In addition to the electrostatic Coulombic potentials, 
molecular Hamiltonians have a lot of microscopic terms 
which explain relativistic effects, fine and hyperfine structure, 
e.g., there are 27 terms derived for general diatomic molecules in \cite{BrownCarrington_2003} 
including electronic spin-spin, spin-orbit couplings and electronic spin-nuclear spin coupling. 
Such microscopic interactions give energy scales comparable to 
thermal energies in cold and ultracold temperatures ($10\,{\rm mK}-10\,{\rm nK}$), 
while the above electronic and vibrational motions corresponds to 
much higher temperatures, $10^4\,{\rm K}$ and $10^3\,{\rm K}$, respectively.  
Thus for the purpose of cold and ultracold physics, 
it is convenient to derive an {\it effective Hamiltonian} 
which operates only within nuclear rotational, electronic spin, and nuclear spin 
degrees of freedom in the vibrational and electronic ({\it vibronic}) ground state. 
Such an effective Hamiltonian is obtained by including the effects of 
off-diagonal matrix elements in the original Hamiltonian 
which couple the vibronic ground state to other excited states. 
Suppose that our full Hamiltonian can be divided into 
a dominant part $\hat{H}_0$ and perturbative part $\lambda\hat{H}_1$ 
with a perturbation parameter $\lambda$, 
\beq
\hat{H}&=&\hat{H}_0+\lambda\hat{H}_1~.
\eeq
In the molecule, $\hat{H}_0$ is composed of the non-relativistic kinetic energy operators of electrons, 
the vibrational terms of nuclei, and the electrostatic Coulombic potentials, 
while $\lambda\hat{H}_1$ is composed of all the remaining terms, 
nuclear rotation, electronic spin-spin, spin-orbit couplings and electronic spin-nuclear spin coupling, and so on. 
The eigenstates of $\hat{H}_0$ are purely vibronic eigenstates, 
\beq
\hat{H}_0\ket{\eta;k}_0&=&E^{0}_{\eta}\ket{\eta;k}_0
~, 
\eeq
where $\eta$ denotes the vibronic quantum numbers and $k$ represents 
nuclear rotational, electronic spin, and nuclear spin quantum numbers. 
The perturbative term $\lambda\hat{H}_1$ mixes these quantum numbers 
and the eigenstates of $\hat{H}$ are given by 
\beq
\hat{H}\ket{\bar{\eta};\bar{k}}&=&E_{\bar{\eta};\bar{k}}\ket{\bar{\eta};\bar{k}}~.
\eeq
Here we used {\it adiabatic labels} $\bar{\eta}$ and $\bar{k}$ assuming that 
the state $\ket{\bar{\eta};\bar{k}}$ coincides with $\ket{\eta;k}_0$ 
when we adiabatically turn off the perturbative term. 
The effective Hamiltonian $\hat{H}_{\rm eff}$ of our interest 
should act only on a subspace $\mathcal{H}_{\rm eff}$ spanned by the vibronic ground states $\{\ket{0;k}_0\}_k$ of $\hat{H}_0$, 
while reproducing the eigenenergies $\{E_{\bar{0};\bar{k}}\}_{\bar{k}}$ of the full Hamiltonian. 
Therefore, the effective Hamiltonian should satisfy an eigenvalue equation such that 
\beq
\hat{H}_{\rm eff}\ket{0;\bar{k}}_{\rm eff}
&=&E_{\bar{0};\bar{k}}\ket{0;\bar{k}}_{\rm eff}~,
\eeq
with eigenstates $\ket{0;\bar{k}}_{\rm eff}$ given by linear combinations of states only in $\mathcal{H}_{\rm eff}$. 
This condition gives a perturbative expansion of $\hat{H}_{\rm eff}$ as 
\beq
\hat{H}_{\rm eff}&=&
\hat{H}_0
+\lambda \hat{P}_0 \hat{H}_1 \hat{P}_0 
+\lambda^2 \hat{P}_0 \hat{H}_1\frac{\hat{Q}_0}{E^0_0-\hat{H}_0}\hat{H}_1 \hat{P}_0 
+\mathcal{O}(\lambda^3)
\label{H_eff}
~,
\eeq
where $\hat{P}_0$ is a projection operator onto the subspace $\mathcal{H}_{\rm eff}$ 
and $\hat{Q}_0$ is a projection operator onto its orthogonal complement, 
\beq
\hat{P}_0&=&\sum_{i} {\mathstrut\ket{0;i}}_0 \, {\mathstrut}_0\!\bra{0;i} ~,\\
\hat{Q}_0&=& {\bf 1}-\hat{P}_0
~=~\sum_{\eta\neq0}\sum_{i}{\mathstrut \ket{\eta;i}}_0 \, {\mathstrut}_0\!\bra{\eta;i}~.
\eeq
A detailed derivation and fuller expression for Eq.~(\ref{H_eff}) can be found in \cite{BrownCarrington_2003}. 
The first and second terms in Eq.~(\ref{H_eff}) are just the direct projection of the full Hamiltonian 
onto the subspace of the vibronic ground states without perturbation. 
The third term in Eq.~(\ref{H_eff}) gives the effects of coupling between 
the vibronic ground and excited states. 
It should be remarked that 
the $\Lambda$-doubling degeneracies are broken ({\it $\mathit{\Lambda}$-doubling splitting}) 
by the admixture of the rotational levels in the $X{}^{2}\Pi$ state 
with the corresponding levels of the $A{}^{2}\Sigma$ state via electronic spin-orbit interaction 
and the nuclear rotational-electronic orbit interaction \cite{Mulliken_1931}. 
This is because each rotational level of the $A{}^{2}\Sigma$ state has a definite parity with respect to spatial inversion, 
either positive or negative, and thus interacts with only one of the two parity states in the $X{}^{2}\Pi$ state. 
Also, the centrifugal distortion corrections are obtained by the admixture of the vibrational ground and excited states. 

Throughout this paper, we will employ the effective Hamiltonian for the OH molecule 
acting on rotational, fine and hyperfine levels in its vibronic ground state  
and investigate its energy levels in the presence of electric and magnetic fields.

\section{Hyperfine structure of OH molecule without external fields}

\begin{figure}[t]
\begin{center}
\includegraphics[width=6.5cm]{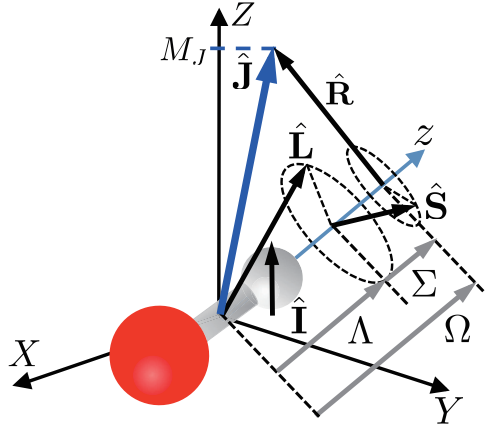}
\end{center}
\vspace{-0.5cm}
\caption{Vector coupling diagram for OH molecule in Hund's case (a) basis 
(red = oxygen, white = hydrogen, grey = idealized bond). 
The space-fixed coordinate system is represented by $XYZ$, 
while the symmetry axis of the molecule is labeled by the molecule-fixed $z$-axis. 
The electronic orbital and spin angular  momenta are represented by $\hat{\bf L}$ and $\hat{\bf S}$, respectively, 
and $\Lambda$ and $\Sigma$ are their projection along the symmetry axis of molecule. 
$\hat{\bf R}$ is the rotational angular momentum of the nuclei, 
$\hat{\bf I}$ is the nuclear spin angular momentum, 
and $\hat{\bf  J}$ is the total angular momentum without the nuclear spin,  
$\hat{\bf  J}=\hat{\bf R}+\hat{\bf L}+\hat{\bf S}$. 
The projection of $\hat{\bf  J}$ along the space-fixed $Z$-axis is given by $M_J$, 
while the projection along the molecule-fixed $z$-axis is given by $\Omega=\Lambda+\Sigma$. 
}
\label{FIG_OH_Hund_a}
\end{figure}

In a study of molecular spectroscopy, 
especially at low temperatures where vibrational degrees of freedom for nuclei are frozen out, 
we often use a quantum rotor model, 
replacing a molecule by a quantum-mechanically rotating rigid body with several angular momenta 
\cite{BrownCarrington_2003,Zare_1988}. 
The quantum rotor model takes into account not only the rotation but also interactions between angular momenta, 
depending on the species of molecule. 
Since the OH molecule in its electronic ground state 
is a diatomic molecule with a nuclear spin of $I=1/2$ and has an electron in the open $2p$-shell, 
there are four main kinds of angular momenta: 
the rotational angular momentum of the nuclei $\hat{\bf R}$, 
the electronic orbital angular momentum $\hat{\bf L}$, 
the electronic spin angular momentum $\hat{\bf S}$, 
and the nuclear spin angular momentum $\hat{\bf I}$, see Figure~\ref{FIG_OH_Hund_a}. 
Furthermore, Furthermore, the total angular momentum without the nuclear spin is conventionally defined as 
$\hat{\bf  J}=\hat{\bf R}+\hat{\bf L}+\hat{\bf S}$, 
and the geometric part of angular momenta as $\hat{\bf N}=\hat{\bf R}+\hat{\bf L}$. 
Here we remark that for the OH molecule in its ${}^2\Pi$ states 
the  magnitudes of the angular momenta $\hat{\bf L}$, $\hat{\bf S}$, and $\hat{\bf I}$ 
are fixed to be $L=1$, $S=1/2$, and $I=1/2$, respectively, 
while the rotational quantum number $R$ takes non-negative integer values. 
In this section, we review the quantum rotor model of the OH molecule in the absence of external fields, 
and then numerically calculate its energy spectrum including hyperfine structure, leading to 
the lowest 16 states and therefore reproducing experimental data to an accuracy of 2\,kHz. 
\ref{sec:stp} summarizes definitions of spherical tensor operators and their formulae, 
which we utilize in calculations of matrix elements throughout this paper. 

\subsection{The zero-field effective Hamiltonian for OH}

\begin{figure}[t]
\begin{center}
\includegraphics[width=14cm]{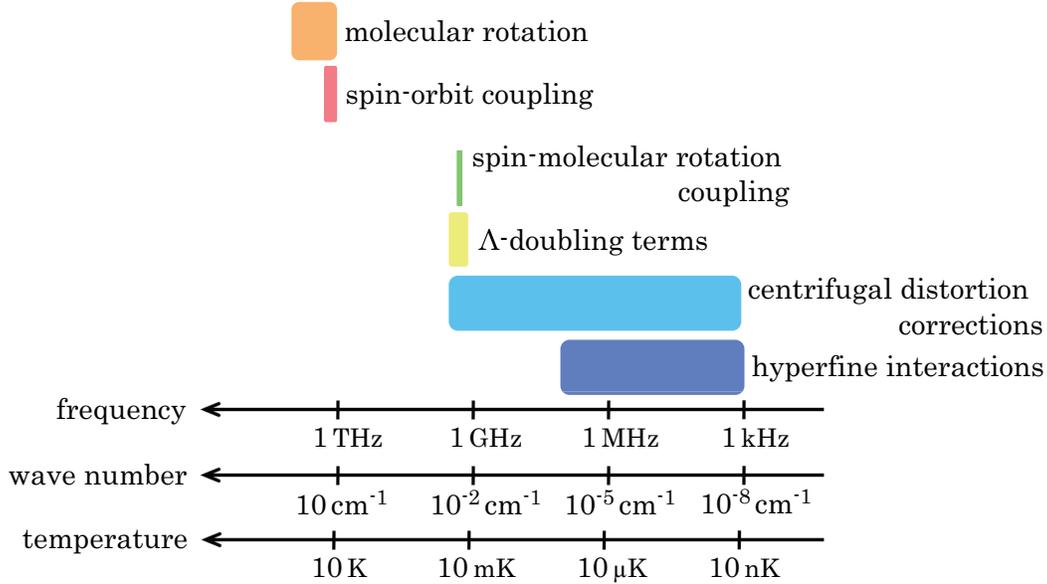}
\end{center}
\vspace{-0.5cm}
\caption{
Energy scales of the effective Hamiltonian for OH in the $v=0$ level of the $X{}^2\Pi$ ground state. 
Each energy scale is estimated by the size of matrix elements in the subspace of the lowest 96 states on a logarithmic scale. 
We consider frequency $\nu$, wavenumber $1/\lambda$, and temperature $T$ as units of energy by 
multiplying them through by the fundamental constants $h\nu$, $hc/\lambda$, and $k_{B}T$, respectively. 
These units of energy relate to each other: $1\,{\rm MHz}\simeq50\,\mu{\rm K}\simeq3.3\times10^{-5}\,{\rm cm}^{-1}$.
}
\label{FIG_energy_scales_H0}
\end{figure}

We first summarize the zero-field effective Hamiltonian for the OH molecule 
in the $v=0$ level of the $X{}^2\Pi$ state. 
Throughout this paper, we use notations in molecular spectroscopy 
to represent the vibrational and electronic ground state of molecules by $v=0$ and $X$, respectively, and 
to classify the molecular states with ${}^{2S+1}\Pi_{|\Lambda+\Sigma|}$ 
where $\Lambda$ and $\Sigma$ are the projections of the electronic orbital and spin angular momenta along 
the molecule-fixed $z$-axis, respectively, and $\Pi$ stands for $|\Lambda|=1$ \cite{BrownCarrington_2003,Zare_1988}. 
The effective Hamiltonian is given by \cite{BrownCarrington_2003,Brown_1978,Brown_1979,Brown_1981}
\beq
\hat{H}_0&=&
\hat{H}_{\rm SO} + \hat{H}_{\rm MR} + \hat{H}_{\rm SMR} + \hat{H}_{\rm LD} + \hat{H}_{\rm HF} + \hat{H}_{\rm CD}  \, ,
\label{H0}
\eeq
where $\hat{H}_{\rm SO}$ represents the spin-orbit coupling, 
\beq
\hat{H}_{\rm SO}&=&
A_{\rm SO}T^1_{q=0}(\hat{\bf L})T^1_{q=0}(\hat{\bf S}) \, ,
\label{HSO}
\eeq
$\hat{H}_{\rm MR}$ represents  the rotational energy of the molecule, 
\beq
\hat{H}_{\rm MR}&=& 
B_{N}\hat{\bf N}^2 \, ,
\label{HMR}
\eeq
$\hat{H}_{\rm SMR}$ is the spin-molecular rotation coupling, 
\beq
\hat{H}_{\rm SMR}&=&
\gamma\, T^1(\hat{\bf  J}-\hat{\bf S})\cdot T^1(\hat{\bf S}) \, ,
\label{HSMR}
\eeq
$\hat{H}_{\rm LD} $ denotes the $\Lambda$-doubling terms, 
\beq
\hat{H}_{\rm LD}&=&
\sum_{q=\pm 1}{\rm e}^{-2{\rm i}q\phi}\,
\left[ -Q\, T^2_{2q}(\hat{\bf  J},\hat{\bf  J})+(P+2Q)T^2_{2q}(\hat{\bf  J},\hat{\bf S}) \right] \, ,
\label{HLD}
\eeq
and $\hat{H}_{\rm HF}$ represents the hyperfine interactions, 
\beq
\fl
\hat{H}_{\rm HF} 
~=~ a\,T^1_{q=0}(\hat{\bf I})T^1_{q=0}(\hat{\bf L}) + b_F T^1(\hat{\bf I})\cdot T^1(\hat{\bf S}) 
+ \sqrt{\frac{2}{3}}c\, T^2_{q=0}(\hat{\bf I},\hat{\bf S}) 
+ d\sum_{q=\pm 1} {\rm e}^{-2{\rm i}q\phi}\, T^2_{2q}(\hat{\bf I},\hat{\bf S}) 
\nonumber\\ 
\hspace{-40pt}
 + c_I T^1(\hat{\bf I})\cdot T^1(\hat{\bf  J}-\hat{\bf S}) 
+ c_I' \sum_{q=\pm 1} {\rm e}^{-2{\rm i}q\phi}\,\frac{1}{2}
\left[ T^2_{2q}(\hat{\bf I},\hat{\bf  J}-\hat{\bf S})+T^2_{2q}(\hat{\bf  J}-\hat{\bf S},\hat{\bf I}) \right] \,.  
\label{HHF}
\eeq
Finally, the centrifugal distortion corrections to the above terms are given by 
\beq
\fl 
\hat{H}_{\rm CD}
~=~
-D(\hat{\bf N}^2)^2+H(\hat{\bf N}^2)^3
+\gamma_D\, \left\{T^1(\hat{\bf  J}-\hat{\bf S})\cdot T^1(\hat{\bf S})\right\}\hat{\bf N}^2
\nonumber\\
\hspace{-40pt}
+\sum_{q=\pm 1}{\rm e}^{-2{\rm i}q\phi}\,
\biggl\{
-\frac{Q_D}{2}
\left[
T^2_{2q}(\hat{\bf  J},\hat{\bf  J})\hat{\bf N}^2+\hat{\bf N}^2T^2_{2q}(\hat{\bf  J},\hat{\bf  J})
\right]
+\frac{P_D+2Q_D}{2}
\left[
T^2_{2q}(\hat{\bf  J},\hat{\bf S})\hat{\bf N}^2+\hat{\bf N}^2T^2_{2q}(\hat{\bf  J},\hat{\bf S})
\right]
\biggl\}
\nonumber\\
\hspace{-40pt}
+\sum_{q=\pm 1}{\rm e}^{-2{\rm i}q\phi}\,
\biggl\{
-\frac{Q_H}{2}
\left[
T^2_{2q}(\hat{\bf  J},\hat{\bf  J})(\hat{\bf N}^2)^2+(\hat{\bf N}^2)^2\,T^2_{2q}(\hat{\bf  J},\hat{\bf  J})
\right]
\nonumber\\
\hspace{35pt}
+\frac{P_H+2Q_H}{2}
\left[
T^2_{2q}(\hat{\bf  J},\hat{\bf S})(\hat{\bf N}^2)^2+(\hat{\bf N}^2)^2\,T^2_{2q}(\hat{\bf  J},\hat{\bf S})
\right]
\biggl\}
\nonumber\\
\hspace{-40pt}
+d_D\sum_{q=\pm 1} {\rm e}^{-2{\rm i}q\phi}\, 
\frac{1}{2}
\left[
T^2_{2q}(\hat{\bf I},\hat{\bf S})\hat{\bf N}^2+\hat{\bf N}^2T^2_{2q}(\hat{\bf I},\hat{\bf S})
\right]
\, .
\label{HCD}
\eeq
\begin{table}[t]
\begin{center}
\caption{\label{Table_para}
Molecular parameters for OH in the $v=0$ level of the $X{}^2\Pi$ ground state (in MHz) \cite{Brown_1981}.
} 
\lineup
\begin{tabular}{@{}*{4}{l}}
\br                              
\qquad$A_{\rm so}\,=\,-4 168 639.13(78)$ \qquad \cr 
\qquad$B_N\,=\,555 660.97(11)$ ~~ \cr 
\qquad$\gamma\,=\,-3574.88(49)$~~ \cr
\qquad$Q\,=\,-1159.991 650$ ~~ \cr 
\qquad$P\,=\,7053.098 46$ ~~ \cr 
\qquad$a\,=\,86.1116$ ~~ \cr 
\qquad$b_F\,=\,-73.2537$ ~~  \cr
\qquad$c\,=\,130.641$ ~~ \cr 
\qquad$d\,=\,56.6838$ ~~ \cr
\qquad$c_I\,=\,-0.099 71$ ~~ \cr 
\qquad$c_I'\,=\,0.643\times 10^{-2}$ ~~ \cr
\qquad$D\,=\,57.1785(86)$ ~~ \cr 
\qquad$H\,=\,0.4236\times 10^{-2}$ ~~ \cr 
\qquad$\gamma_D\,=\,0.7315$ ~~ \cr
\qquad$Q_D\,=\,0.4420320$ ~~ \cr 
\qquad$P_D\,=\,-1.550 962$ ~~ \cr
\qquad$Q_H\,=\,-0.8237\times 10^{-4}$ ~~ \cr 
\qquad$P_H\,=\,0.1647\times 10^{-3}$ ~~ \cr 
\qquad$d_D\,=\,-0.02276$~ \cr
\br
\end{tabular}
\end{center}
\end{table}
Note that 
these operators are written for calculation of their matrix elements with use of a Hund's case (a) basis in the molecule-fixed frame 
\cite{BrownCarrington_2003,Zare_1988}, 
and that $q$ denotes the component of the spherical tensor operator 
and $\phi$ is the azimuthal coordinate of the electronic orbit in the molecule-fixed frame. 
For the OH molecule, the Hund's case (a) basis consists of the simultaneous eigenstates of angular momenta, 
$\hat{\bf L}^2$, $\hat{\bf S}^2$, $\hat{\bf  J}^2$ and $\hat{\bf I}^2$, 
 projections on the molecule-fixed $z$-axis, $\hat{L}_z$, $\hat{S}_z$ and $\hat{J}_z$, 
and projections on the space-fixed $Z$-axis, $\hat{J}_Z$ and $\hat{I}_Z$, 
specified by $\ket{L\Lambda,\! S\Sigma, \! J\Omega M_J, \! IM_I}$ 
with $L=1$, $\Lambda=\pm1$, $S=1/2$, $\Sigma=\pm1/2$, 
$J\geq1/2$, $\Omega=\Lambda+\Sigma$, $-J\leq M_J\leq J$, $I=1/2$, and $M_I=\pm1/2$. 
The matrix elements in the Hund's case (a) basis are given in \ref{sec:matele}. 
Table~\ref{Table_para} lists the values of molecular parameters of Eqs.\,(\ref{HSO})-(\ref{HCD}), 
extracted from \cite{Brown_1981}, and Figure~\ref{FIG_energy_scales_H0} shows 
the energy scales of the effective Hamiltonian for OH. 
It is clear that one must take into account 
the centrifugal distortion corrections and the hyperfine interactions 
in order to obtain access to and investigate the physics of OH molecules at microKelvin temperatures and below. 


\subsection{The zero-field energy spectrum}

We will investigate the energy spectrum of the Hamiltonian Eq.\,(\ref{H0}), 
focusing on the hyperfine structure of the lowest 16 states in the ${}^2\Pi_{3/2}$ manifold. 
Before going into details of the spectrum, let us overview the hierarchy of energy scales in the OH molecule. 
First of all, for the OH molecule the spin-orbit coupling gives the largest energy scale $A_{\rm so}\sim -4$\,THz. 
Noting that the matrix elements of the spin-orbit coupling in Eq.\,(\ref{HSO}) are diagonal in the Hund's case (a) basis 
and proportional to $\Lambda\Sigma$ (see \ref{sec:matele}), 
the ${}^2\Pi$ states can be classified in two cases: one with $\Lambda\Sigma=1/2$, and 
the other with $\Lambda\Sigma=-1/2$. 
The former case has the lower energy and it is also labeled by $|\Lambda+\Sigma|=3/2$, called the ${}^2\Pi_{3/2}$ manifold, 
while the latter is labeled by $|\Lambda+\Sigma|=1/2$, corresponding to the ${}^2\Pi_{1/2}$ manifold. 
The next largest energy scale is given by the rotational energy of molecule in Eq.\,(\ref{HMR}) with $B_{N}\sim 0.5$\,THz. 
Since in the ${}^2\Pi_{3/2}$ manifold the molecular-axis projections $\Lambda$ and $\Sigma$ point in the same direction, 
the total angular momentum starts from $J=3/2$ and increases by positive integers as the nuclei  rotate faster. 
On the other hand, in the ${}^2\Pi_{1/2}$ manifold $\Lambda$ and $\Sigma$ point in opposite directions, 
and thus the total angular momentum can take the minimum, $J=1/2$. 
Third, the spin-molecular rotation coupling in Eq.\,(\ref{HSMR}), where $\gamma\sim-3$\,GHz, 
becomes diagonal in our basis if we only consider 
the lowest 24 states consisting of $J=3/2$ states in the ${}^2\Pi_{3/2}$ manifold and $J=1/2$ states in the ${}^2\Pi_{1/2}$ manifold. 
However, there appear off-diagonal matrix elements that contribute 
when we take into account higher rotational states, and have to be folded in as corrections. 
Fourth, the $\Lambda$-doubling terms Eq.\,(\ref{HLD}) with $Q\sim-1$\,GHz and $P\sim 7$\,GHz yield 
the $\Lambda$-doubling splittings in the spectrum, 
especially assigning 1\,GHz splitting in the lowest $J=3/2$ states of the ${}^2\Pi_{3/2}$ manifold. 
We remark that to obtain the $\Lambda$-doubling splittings in the $J=3/2$ states of ${}^2\Pi_{3/2}$ 
we must include at least the $J=3/2$ states of ${}^2\Pi_{1/2}$ in our model. 
Note also that  the $\Lambda$-doubling terms hybridize fixed-$\Omega$ states, 
yielding the parity-conserved states in ${}^2\Pi_{3/2}$ as an appropriate basis in the absence of external fields, 
\beq\fl
\lefteqn{
\ket{L, S ;  J,|\Omega|, M_J ; I,M_I; \epsilon\,}
}\nonumber\\
\hspace{-70pt}
=~
\bigl\{\,
\ket{L,|\Lambda|}\ket{S,|\Sigma|} \ket{J,|\Omega|, M_J}\ket{I,M_I}
+\epsilon(-1)^{J-S}\ket{L,-|\Lambda|}\ket{S,-|\Sigma|} \ket{J,-|\Omega|, M_J}\ket{ I,M_I}
\bigl\}/\sqrt{2}
\, .
\label{parity_state}
\eeq
Here $\epsilon$ takes values of $1$ and $-1$ corresponding to the positive and negative parity states, respectively. 
Finally, hyperfine interactions Eq.\,(\ref{HHF}) and centrifugal distortion effects Eq.\,(\ref{HCD}) 
give further microscopic structure in energy of tens of megahertz. 
As we will see later, the hyperfine interactions play a significant role in 
the emergence of level repulsions when we apply electric and magnetic fields. 

\begin{figure}[t]
\begin{center}
\includegraphics[width=14cm]{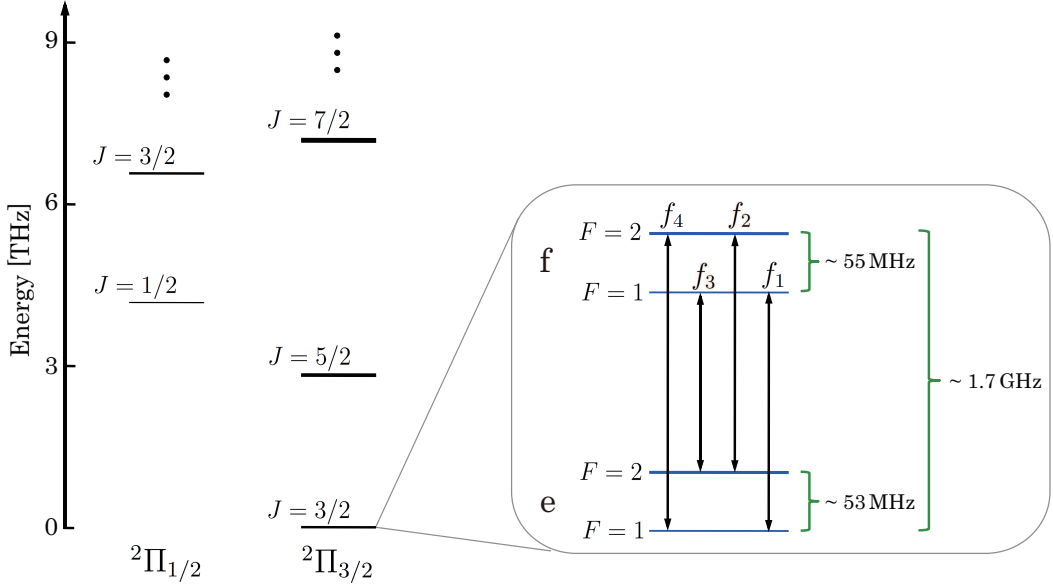}
\end{center}
\vspace{-0.5cm}
\caption{Zero-field energy spectrum and hyperfine structure of OH in the $v=0$, $X{}^2\Pi_{3/2}$ ground state. 
Each rotational level, labeled by $J$, contains $4(2J+1)$ states. 
A closeup of the hyperfine structure in $J=3/2$ states of ${}^2\Pi_{3/2}$ shows the $\Lambda$-doubling splittings, 
represented by ``e"-states for the negative parity states and ``f"-states for the positive parity states, 
and also hyperfine splittings with transition frequencies $f_1$, $f_2$, $f_3$, and $f_4$. 
}
\label{FIG_spec_H0}
\end{figure}

We proceed to determine the energy spectrum of the Hamiltonian Eq.\,(\ref{H0}) numerically. 
Since we are interested in the spectrum of the lowest 16 states in the ${}^2\Pi_{3/2}$ manifold, 
we employed cutoffs for the unbounded quantum number $J$ 
to consider a finite-dimensional subspace of states. 
More precisely, we restricted ourselves to states with $J=1/2$ (8 states) and $J=3/2$ (16 states) in the ${}^2\Pi_{1/2}$ manifold, 
and $J=3/2$ (16 states), $J=5/2$ (24 states), and $J=7/2$ (32 states)  in the ${}^2\Pi_{3/2}$ manifold, 
that is, the lowest 96 states as shown in Figure~\ref{FIG_spec_H0}. 
Note that each rotational level labeled by $J$ has further internal degrees of freedom, i.e., 
signs of $\Lambda$ ($\Lambda=\pm|\Lambda|$), projections of $\hat{\bf  J}$ ($M_J=0,\pm1,\dots,\pm J$), 
and projections of the nuclear spin ($M_I=\pm 1/2$), 
thus it contains $4(2J+1)$ states. 
Figure~\ref{FIG_spec_H0} also shows a closeup of the hyperfine structure in $J=3/2$ states of ${}^2\Pi_{3/2}$ 
where we described the $\Lambda$-doubling splittings 
by ``e"-states for the negative parity states and ``f"-states for the positive parity states, 
and hyperfine splittings with transition frequencies $f_1$, $f_2$, $f_3$, and $f_4$. 
We introduced the total angular momentum $\hat{\bf F}=\hat{\bf  J}+\hat{\bf I}$ 
to label the hyperfine states. 
We numerically diagonalized the zero-field Hamiltonian Eq.\,(\ref{H0}) in the Hund's case (a) basis consisted of the lowest 96 states, 
and calculated the energy spectrum of the lowest 16 states in the ${}^2\Pi_{3/2}$ manifold. 
Table~\ref{Table_H0HF} summarizes our numerical results along with experimental data \cite{Meulen_1972,Ye_2006,Lev_2006}. 
The largest deviation from the experimental data is in our calculation of $f_4$, 
which determines an accuracy of calculations down to $1.8\,{\rm kHz}\simeq86\,{\rm nK}$. 
From our results, we determine the energy of the $\Lambda$-doubling splittings to be 
$\Delta_{\rm LD}=(f_1+f_2+f_3+f_4)/4=1666.3791\,{\rm MHz}$, and
that of the hyperfine splitting for the even-parity states to be 
$\Delta^{\rm (e)}_{\rm HF}=(f_1-f_3+f_4-f_2)/2=53.1706\,{\rm MHz}$ and 
for the odd-parity states to be 
$\Delta^{\rm (f)}_{\rm HF}=(f_2-f_3+f_4-f_1)/2=55.1276\,{\rm MHz}$, respectively. 


\begin{table}
\begin{center}
\caption{\label{Table_H0HF}
Hyperfine transition frequencies (in MHz) for OH in the $v=0$, $X{}^2\Pi_{3/2}$ ground state.} 
\lineup
\begin{tabular}{@{}*{5}{l}}
\br              
~~  &~ Calculated &~ Meulen {\it et al.} \cite{Meulen_1972} &~ Hudson {\it et al.} \cite{Ye_2006}&~ Lev {\it et al.} \cite{Lev_2006} ~~\\
\mr
~~ $f_1$ &~ 1665.4006 &~  1665.40184(10) &~1665.401803(12) &~ \qquad------ ~~\\
~~ $f_2$ &~ 1667.3576 &~ 1667.35903(10) &~ 1667.358996(4) &~ \qquad------ ~~\\
~~ $f_3$ &~ 1612.2300 &~ 1612.23101(20) &~ \qquad------ &~ 1612.230825(15)~~\\
~~ $f_4$ &~ 1720.5282 &~ 1720.52998(10) &~ \qquad------ &~ 1720.529887(10)~~\\           
\br
\end{tabular}
\end{center}
\end{table}

\section{Hyperfine structure of OH molecule in combined electric and magnetic fields}

In this section, we will add the Stark and Zeeman Hamiltonians to the zero-field Hamiltonian Eq.\,(\ref{H0}) 
in order to study the energy spectra under combined electric and magnetic fields. 
We first examine the Stark and Zeeman effects for the spectrum of $J=3/2$ states in the ${}^2\Pi_{3/2}$ manifold separately, 
and then investigate the energy spectra in the presence of combined electric and magnetic fields.

\subsection{The effective Hamiltonian of OH molecule in combined electric and magnetic fields}

\begin{figure}[t]
\begin{center}
\includegraphics[width=6cm]{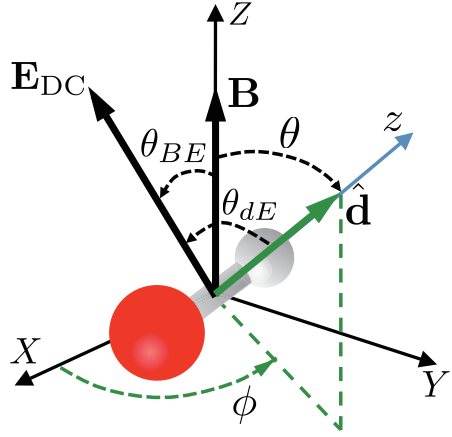}
\vspace{-0.5cm}
\caption{Field configuration and Euler angles defining a general orientation of the molecule-fixed axes. 
The magnetic field ${\bf B}$ is chosen parallel to the space-fixed $Z$-axis, and the electric DC field ${\bf E}_{\rm DC}$ 
is in the space-fixed $XZ$-plane making an angle of $\theta_{BE}$ with the magnetic field. 
The electric dipole moment operator $\hat{\bf d}$ is parallel to the symmetry axis of molecule and 
$\theta_{dE}$ is the angle between $\hat{\bf d}$ and ${\bf E}_{\rm DC}$. 
The Euler angles are labeled by $\omega=(\phi,\theta,\chi)$, though the angle $\chi$ is not shown here for simplicity. 
}
\label{FIG_OH_EB}
\end{center}
\end{figure}

Let us consider the quantum rotor model for the OH molecule 
in the presence of combined electric and magnetic fields. 
Figure~\ref{FIG_OH_EB} shows a field configuration of electric and magnetic fields 
in which the magnetic field ${\bf B}$ is chosen parallel to the space-fixed $Z$-axis, and the electric DC field ${\bf E}_{\rm DC}$ 
is in the space-fixed $XZ$-plane making an angle of $\theta_{BE}$ with the magnetic field. 
We also introduced Euler angles $\omega=(\phi,\theta,\chi)$ which define a general orientation of the molecule-fixed frame. 
The effective Hamiltonian of the OH molecule 
in the presence of combined electric and magnetic fields is then given by
\beq
\hat{H}
&=& \hat{H}_0 + \hat{H}_{\rm S} + \hat{H}_{\rm Z} \, .    
\label{H_EB}
\eeq
Here  the zero-field Hamiltonian $\hat{H}_0$ is  given in Eq.\,(\ref{H0}). 
The Stark Hamiltonian $\hat{H}_{\rm S}$ is given by 
\beq
\hat{H}_{\rm S} 
&=& -\hat{\bf d}\cdot{\bf E}_{\rm DC} \, ,
\label{HS1}
\eeq 
where $\hat{\bf d}$ is the electric dipole moment operator of OH, having a non-zero component 
along the molecule-fixed $z$-direction, 
\beq
T^1_q(\hat{\bf d})
&=&
\mu^{(e)}_z \delta_{q,0}
\, ,
\eeq
with its permanent electric dipole moment $\mu^{(e)}_z =1.65520(10)$\,Debye \cite{Peterson_1984}. 
Noting that the angle between $\hat{\bf d}$ and ${\bf E}_{\rm DC}$, $\theta_{dE}$, satisfies 
$\cos{\theta_{dE}}=\cos{\theta_{BE}}\cos\theta+\sin{\theta_{BE}}\sin\theta\cos\phi$, 
we can rewrite the Stark Hamiltonian Eq.\,(\ref{HS1}) as
\beq
\hat{H}_{\rm S} 
&=& -\mu^{(e)}_z E_{\rm DC}
 \sum_{p=0, \pm1}
d^{(1)}_{p,0}(\theta_{BE})\mathscr{D}^{(1)\ast}_{p,0}(\omega)
\label{HS2}
\, , 
\eeq
with the matrix elements of the Wigner $D$-matrix $\mathscr{D}^{(J)}_{p,q}$ 
and those of the reduced rotation matrix $d^{(1)}_{p,q}$ \cite{BrownCarrington_2003,Zare_1988}. 
In (\ref{H_EB}), 
the Zeeman Hamiltonian of the OH molecule is defined as 
\beq
\fl
\hat{H}_{\rm Z}
~=~
g_L'\mu_B B_{Z} T^1_{p=0}(\hat{\bf L})
+g_S\mu_B B_{Z} T^1_{p=0}(\hat{\bf S})
-g_r\mu_B B_{Z} T^1_{p=0}(\hat{\bf  J}-\hat{\bf L}-\hat{\bf S})
-g_N\mu_N  B_{Z} T^1_{p=0}(\hat{\bf I})
\nonumber\\
\hspace{-40pt}
+g_l\mu_B B_{Z} \sum_{q=\pm1}\mathscr{D}_{0,q}^{(1)\ast}(\omega)T^1_{q}(\hat{\bf S})
+g_l'\mu_B B_{Z} \sum_{q=\pm1}
{\rm e}^{-2{\rm i}q\phi}\,
\mathscr{D}_{0,-q}^{(1)\ast}(\omega)T^1_{q}(\hat{\bf S})
\nonumber\\
\hspace{-40pt}
-g_r^{e'}\mu_B B_{Z} \sum_{q=\pm1}\sum_{p=0,\pm1}
{\rm e}^{-2{\rm i}q\phi}\,
(-1)^p\,
\mathscr{D}_{-p,-q}^{(1)\ast}(\omega) T^1_{p}(\hat{\bf  J}-\hat{\bf S})
\mathscr{D}_{0,-q}^{(1)\ast}(\omega)
\, ,
\label{HZ}
\eeq
which contains, in order, the electronic orbital Zeeman effect, the electronic spin isotropic Zeeman effect, 
the rotational Zeeman effect, the nuclear spin Zeeman effect, and the electronic spin anisotropic Zeeman effect. 
Finally, the last two terms in Eq.\,(\ref{HZ}) are parity-dependent and non-cylindrical Zeeman effects. 
The $g$-factors for OH in the $X{}^2\Pi$ state are listed in Table~\ref{gfactors}. 

\begin{table}[t]
\begin{center}
\caption{\label{gfactors}
$g$-factors for OH in the $X{}^2\Pi$ state \cite{Brown_1978}.} 
\lineup
\begin{tabular}{@{}*{3}{l}}
\br                              
\qquad$g_L'=1.001 07 (15)$ ~~ \cr 
\qquad$g_S=2.001 52 (36)$ ~~ \cr 
\qquad$g_r=-0.633 (19)\times 10^{-3}$ ~~ \cr
\qquad$g_{\ell}=4.00 (56)\times 10^{-3}$ ~~ \cr 
\qquad$g_{\ell}'=6.386 (30)\times 10^{-3}$ ~~ \cr 
\qquad$g_r^{e'}=2.0446 (23)\times 10^{-3}$ ~~ \cr 
\br
\end{tabular}
\end{center}
\end{table}

\subsection{The Stark effect}

\begin{figure}[t]
\begin{center}
\includegraphics[width=15cm]{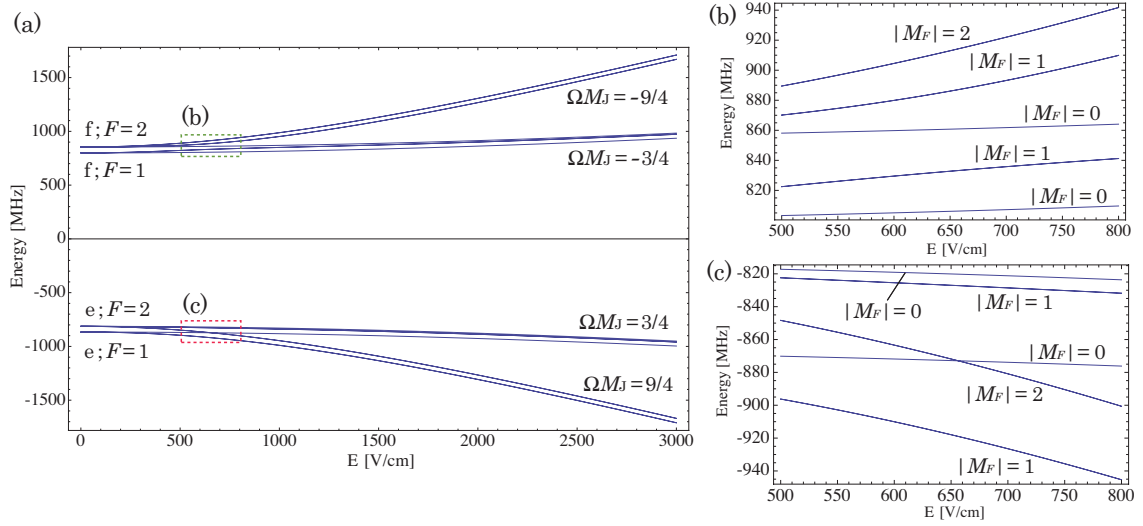}
\end{center}
\vspace{-0.5cm}
\caption{(a) The Stark effect of OH in the $v=0$, $X{}^2\Pi_{3/2}$, $J=3/2$ ground state, 
and the closeups of hyperfine structure: (b) for odd parity states (f-states) in zero field, and (c) for even parity states (e-states). 
There are always two-fold degeneracies with $M_F=\pm |M_F|$ except for $M_F=0$ in the Stark spectrum. 
}
\label{FIG_Stark3000}
\end{figure}

We first examine how the Stark effect modifies the zero electric-field energy spectrum of OH 
in the absence of the magnetic field; $B_{Z}=0$ and $\theta_{BE}=0^{\circ}$. 
As in the calculation of the zero-field energy spectrum, 
we restricted ourselves to the lowest 96 states of OH, 
and diagonalize the Hamiltonian $\hat{H}_0+\hat{H}_{\rm S} $, given in Eqs.~(\ref{H0}) and (\ref{HS2}), 
to obtain the Stark effect of the $v=0$, $X{}^2\Pi_{3/2}$, $J=3/2$ ground state. 
Since in the electronic and vibrational ground states 
electric dipole moments of molecules depend only on their rotational structures, 
their matrix elements can be specified by quantum numbers $J$, $\Omega$, and $M_J$ in the Hund's case (a) basis, 
\beq\fl
\lefteqn{
\bra{L\Lambda' \!,\! S\Sigma' \!,\! J'\Omega' M'_{J'} ,\! IM'_I}
\hat{d}_Z
\ket{L\Lambda,\! S\Sigma, \! J\Omega M_J, \! IM_I}
}
\nonumber\\
\hspace{-70pt}
=
\mu^{(e)}_z
\delta_{\Lambda,\Lambda'}
\delta_{\Sigma,\Sigma'}
\delta_{M_I,M'_{I}}
(-1)^{M_{J'}'-\Omega'}
\sqrt{(2J'+1)(2J+1)}
\biggl(\begin{array}{ccc}
J' & 1 & J \\
-\Omega' & 0 & \Omega
\end{array}\biggl)
\biggl(\begin{array}{ccc}
J' & 1 & J \\
-M_{J'}' & 0 & M_{J}
\end{array}\biggl)
\, .
\eeq
Thus, the diagonal matrix elements of the Stark Hamiltonian $\hat{H}_{\rm S}$ 
with the electric field along the space-fixed $Z$-direction become 
\beq
\bra{L\Lambda,\! S\Sigma, \! J\Omega M_J, \! IM_I}
\hat{H}_{\rm S}
\ket{L\Lambda,\! S\Sigma, \! J\Omega M_J, \! IM_I}
&=&-\mu^{(e)}_z E_{\rm DC}\frac{\Omega M_J}{J(J+1)}
\, ,
\label{lin_S}
\eeq
yielding a linear Stark effect, that is, a first order energy shift in the applied electric field. 
However, due to the $\Lambda$-doubling terms, 
the Hund's case (a) basis states are not the eigenstates of the zero-field Hamiltonian of OH. 
Instead, we can take the parity-conserved states Eq.\,(\ref{parity_state}) 
as an appropriate basis for zero or small applied fields. 
In this basis, the matrix elements become
\beq
\bra{L,\! S, \! J|\Omega| M_J, \! IM_I;\epsilon'}
\hat{H}_{\rm S}
\ket{L,\! S, \! J|\Omega| M_J, \! IM_I;\epsilon}
&=&-\mu^{(e)}_z E_{\rm DC}\frac{|\Omega| M_J}{J(J+1)}
\biggl(\frac{1-\epsilon\epsilon'}{2}\biggl)
\, ,
\label{par_S}
\eeq
showing that the diagonal matrix elements ($\epsilon =\epsilon'$) vanish 
and that the Stark shift is quadratic, i.e., second order in the applied electric field. 
This comes from the fact that 
the electric field is a vector field, a vector-valued function which is odd under spatial inversion, 
and thus the electric field mixes parity states \cite{Freed_1966,Bohn_2002}. 
The structure of the matrix elements Eqs.\,(\ref{lin_S}) and (\ref{par_S}) 
implies that states with different parities repel each other as the electric field is increased, 
and eventually form straight lines with opposite signs in their slopes. 
Figure~\ref{FIG_Stark3000} shows the Stark spectrum of OH in the $v=0$, $X{}^2\Pi_{3/2}$, $J=3/2$ ground state. 
Balancing one half of the $\Lambda$-doubling splitting $\Delta_{\rm LD}$ 
with the linear Stark effect Eq.\,(\ref{lin_S}) averaged over $\Omega M_J=3/4$ and 9/4, 
we can estimate the critical value of the electric field $E_c=2499.84\,{\rm V/cm}\sim 2.5\,{\rm kV/cm}$\footnote{
If we define the critical value of the electric field just by $\mu^{(e)}_z E_c=\Delta_{\rm LD}/2$, we obtain 
$E_c=999.935\,{\rm V/cm}\sim 1\,{\rm kV/cm}$ \cite{Bohn_2002}. 
This definition is simpler than ours but seems inconsistent because 
it takes into account algebraic factors for the $\Lambda$-doubling splitting while not for the Stark effect. }, 
which determines whether the Stark effect becomes linear or quadratic at a given electric field. 
For weak electric fields $ E_{\rm DC}<E_c$, 
the zero-field Hamiltonian $\hat{H}_{0}$ dominates over the Stark term $\hat{H}_{\rm S}$, 
and its perturbative effect based on the parity-conserved states gives 
quadratic curves as shown in Figure~\ref{FIG_Stark3000}(a). 
On the other hand, for strong electric fields $ E_{\rm DC}>E_c$, 
the Stark term becomes dominant, 
and the energy spectrum is well approximated by the linear Stark effect Eq.\,(\ref{lin_S}). 
Figures~\ref{FIG_Stark3000}(b) and (c) show the hyperfine structure in the presence of the electric field 
for states adiabatically connecting to odd and even parity states in zero field, respectively. 
Note that since the projection of the total angular momentum, $M_F$, 
is a good quantum number even in the presence of the electric field \cite{Freed_1966}, 
there are always two-fold degeneracies with $M_F=\pm |M_F|$ except for $M_F=0$. 


\subsection{The Zeeman effect}

\begin{figure}[t]
\begin{center}
\includegraphics[width=12cm]{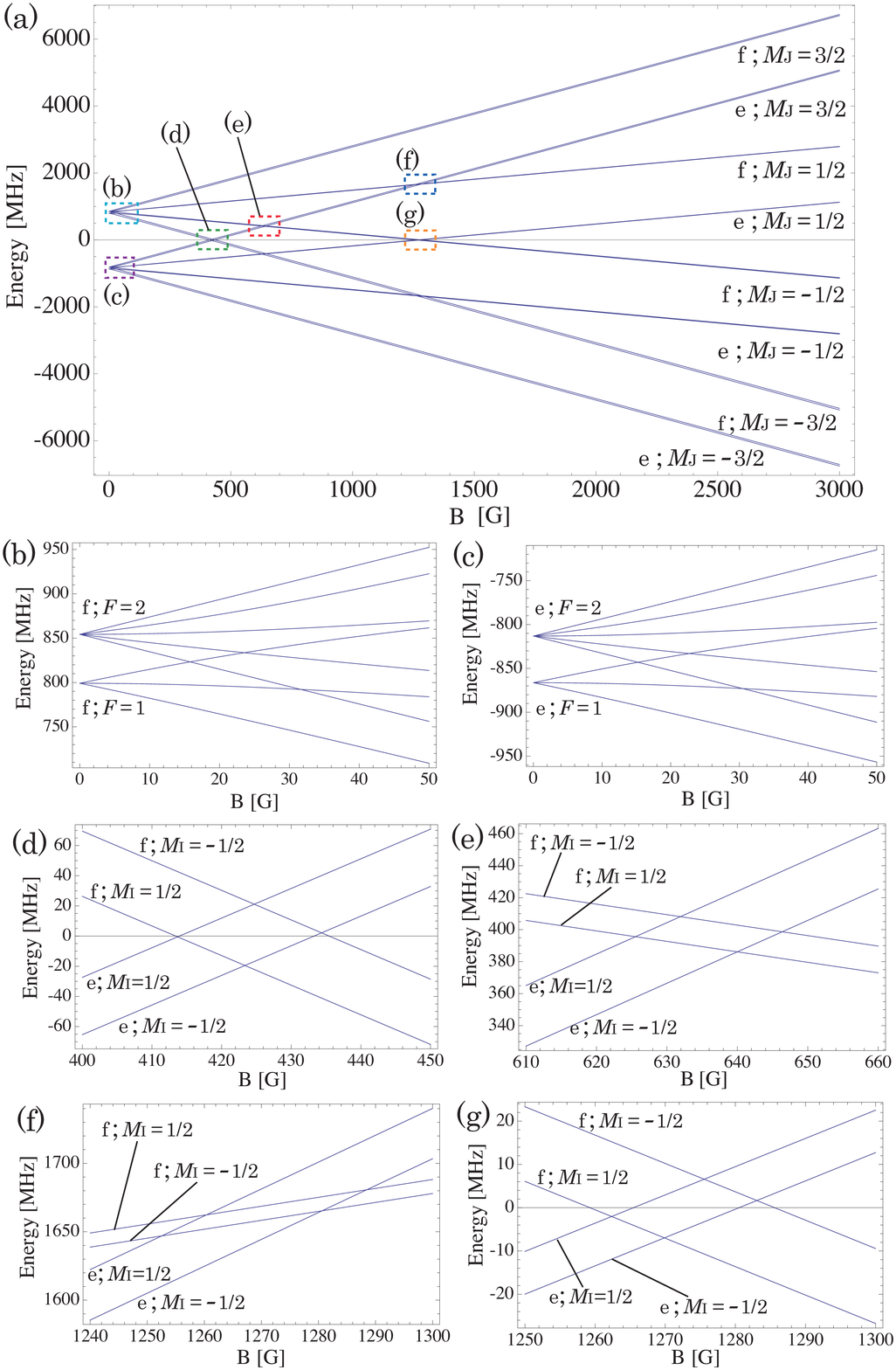}
\end{center}
\vspace{-0.5cm}
\caption{(a) The Zeeman effect for OH in the $v=0$, $X{}^2\Pi_{3/2}$, $J=3/2$ ground state. 
The closeups of hyperfine structure (b) for odd parity states (f-states) in zero field and (c) for even parity states (e-states) 
almost coincide with each other, except for the $\Lambda$-doubling splitting, 
since the dominant part of the Zeeman effect Eq.\,(\ref{par_Z}) is the same. 
(d)-(g) show the level crossings between opposite parity states $\ket{{\rm e};M_I=\pm1/2}$ and $\ket{{\rm f};M_I=\pm1/2}$, 
approaching to the asymptotic states 
(d) $\ket{{\rm e};M_J=3/2}$ and $\ket{{\rm f};M_J=-3/2}$, 
(e) $\ket{{\rm e};M_J=3/2}$ and $\ket{{\rm f};M_J=-1/2}$,
(f) $\ket{{\rm e};M_J=3/2}$ and $\ket{{\rm f};M_J=1/2}$, 
and 
(g) $\ket{{\rm e};M_J=1/2}$ and $\ket{{\rm f};M_J=-1/2}$, 
respectively. 
}
\label{FIG_Zeeman3000}
\end{figure}

We proceed to consider the effect of non-zero static magnetic fields 
in the absence of the electric field; $E_{\rm DC}=0$. 
As before, we treat only the lowest 96 states of OH 
and diagonalize the Hamiltonian $\hat{H}_0+\hat{H}_{\rm Z} $, given in Eqs.\,(\ref{H0}) and (\ref{HZ}), 
to study the Zeeman effect of the $v=0$, $X{}^2\Pi_{3/2}$, $J=3/2$ ground state. 
The Zeeman Hamiltonian of OH is dominated by contributions from 
the electronic orbital and spin isotropic Zeeman effects, as is seen from Table~\ref{gfactors}. 
In the Hund's case (a) basis, the diagonal matrix elements of these two operators 
can be written as 
\beq
\lefteqn{
\bra{L\Lambda,\! S\Sigma, \! J\Omega M_J, \! IM_I}
\bigl\{
g_L'\mu_B B_{Z} T^1_{p=0}(\hat{\bf L})
+g_S\mu_B B_{Z} T^1_{p=0}(\hat{\bf S})
\bigl\}
\ket{L\Lambda,\! S\Sigma, \! J\Omega M_J, \! IM_I}
}
\nonumber\\
&=&
\mu_B B_{Z}(g_L'\Lambda+g_S\Sigma)\frac{\Omega M_J}{J(J+1)}
\, ,
\label{lin_Z}
\eeq
which become in the parity-conserved basis of the ${}^2\Pi_{3/2}$ states
\beq
\lefteqn{
\bra{L,\! S, \! J|\Omega| M_J, \! IM_I;\epsilon'}
\bigl\{
g_L'\mu_B B_{Z} T^1_{p=0}(\hat{\bf L})
+g_S\mu_B B_{Z} T^1_{p=0}(\hat{\bf S})
\bigl\}
\ket{L,\! S, \! J|\Omega| M_J, \! IM_I;\epsilon}
}
\nonumber\\
&=&
\mu_B B_{Z} 
(g_L'|\Lambda|+g_S|\Sigma|)\frac{|\Omega| M_J}{J(J+1)}
\biggl(\frac{1+\epsilon\epsilon'}{2}\biggl)
\, . 
\label{par_Z}
\eeq
Note here that the magnetic field is a pseudovector field, 
a vector-valued function which is even under spatial inversion, 
and therefore the magnetic field does not mix parity states 
while the electric field does \cite{Freed_1966,Bohn_2002}. 
This can be seen in the matrix elements Eq.\,(\ref{par_Z}) which are diagonal in the parity-conserved basis. 
We illustrate this behavior in our Zeeman spectrum Figure~\ref{FIG_Zeeman3000}(a) 
with closeups shown in Figures~\ref{FIG_Zeeman3000}(b) and (c). 
The Zeeman spectrum of OH in the $v=0$, $X{}^2\Pi_{3/2}$, $J=3/2$ ground state 
is composed of two spectra with parity states opposite to each other. 
The $\Lambda$-doubling splitting separates these two spectra by $\Delta_{\rm LD}$, 
but except for such splitting they almost coincide with each other since the dominant part of the Zeeman effect Eq.\,(\ref{par_Z}) 
is the same for both the e- and f-states. 
Noting that $M_F$ is a good quantum number in the absence of applied fields, 
we define a scale of the magnetic field that changes the good quantum number from $M_F$ to $M_J$. 
We set  the hyperfine splittings $\Delta^{\rm (e)}_{\rm HF}$ and $\Delta^{\rm (f)}_{\rm HF}$ 
equal to the linear Zeeman effect Eq.\,(\ref{par_Z}) averaged over $M_J=1/2$ and 3/2, 
and then obtain the critical values of the magnetic field; 
$B^{\rm (e)}_c=63.2576\,{\rm G}$ for the e-states and $B^{\rm (f)}_c=65.5859\,{\rm G}$ for the f-states. 
For strong magnetic fields $B_{Z}>B^{\rm (e)}_c,B^{\rm (f)}_c$, 
the Zeeman effect becomes dominant over the hyperfine structure in the zero-field, 
and the energy spectrum is well approximated by the linear Zeeman effect Eq.\,(\ref{par_Z}) 
in which each energy level can be specified by the quantum number $M_J$ with hyperfine structure labeled by $M_I$,  
as shown in Figures~\ref{FIG_Zeeman3000}(d)-(g). 
In contrast to the Stark spectrum Figure~\ref{FIG_Stark3000}(a), 
the $\Lambda$-doubling splitting remains relevant over the whole range of the magnetic field. 


\subsection{Energy spectra in combined electric and magnetic fields}

\begin{figure}[t]
\begin{center}
\includegraphics[width=12cm]{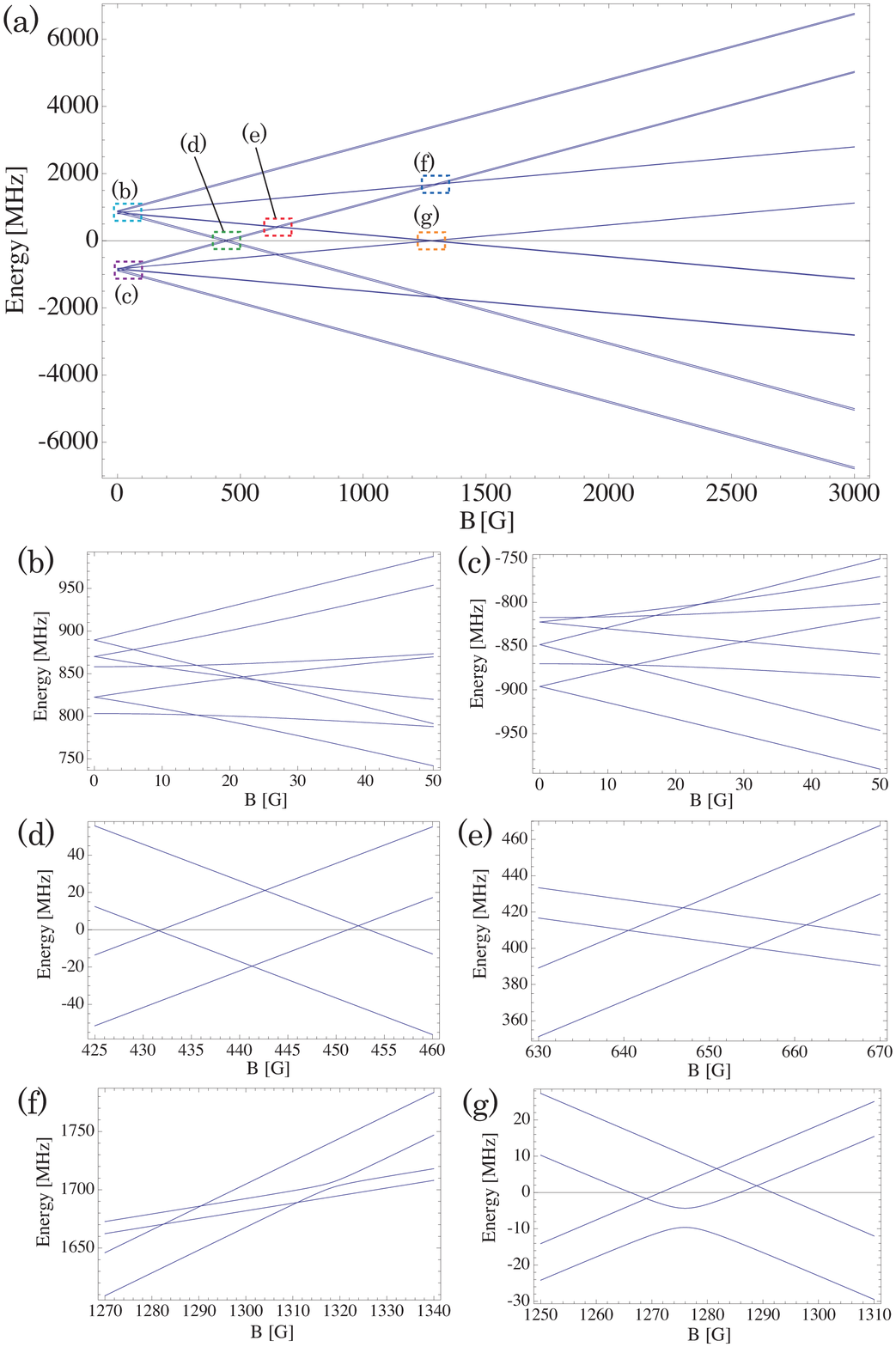}
\end{center}
\vspace{-0.5cm}
\caption{(a) The Zeeman effect for OH in the $v=0$, $X{}^2\Pi_{3/2}$, $J=3/2$ ground state, subject to a bias electric field 
with its strength $E_{\rm DC}=500\,{\rm V/cm}$ and relative angle to the magnetic field $\theta_{BE}=0^{\circ}$. 
(b) The closeups of hyperfine structure for odd parity states (f-states) in zero field and (c) for even parity states (e-states).  
(d)-(g) show the level crossings between opposite parity states $\ket{{\rm e};M_I=\pm1/2}$ and $\ket{{\rm f};M_I=\pm1/2}$, 
similar to Figures~\ref{FIG_Zeeman3000}(d)-(g); however, here level repulsions also appear in (f) and (g) 
due to the Stark effect and hyperfine interactions. 
}
\label{FIG_Zeeman_S00500}
\end{figure}

In the presence of both electric and magnetic fields, 
the energy spectrum of the OH molecule becomes more complicated 
since the Stark effect mixes parity states while the Zeeman effect does not. 
In addition, the hyperfine interactions Eq.\,(\ref{HHF}) give rise to a new kind of level repulsion, 
which we call \textit{Stark-induced hyperfine level repulsion}. 



\begin{figure}[hptb]
\begin{center}
\includegraphics[width=12cm]{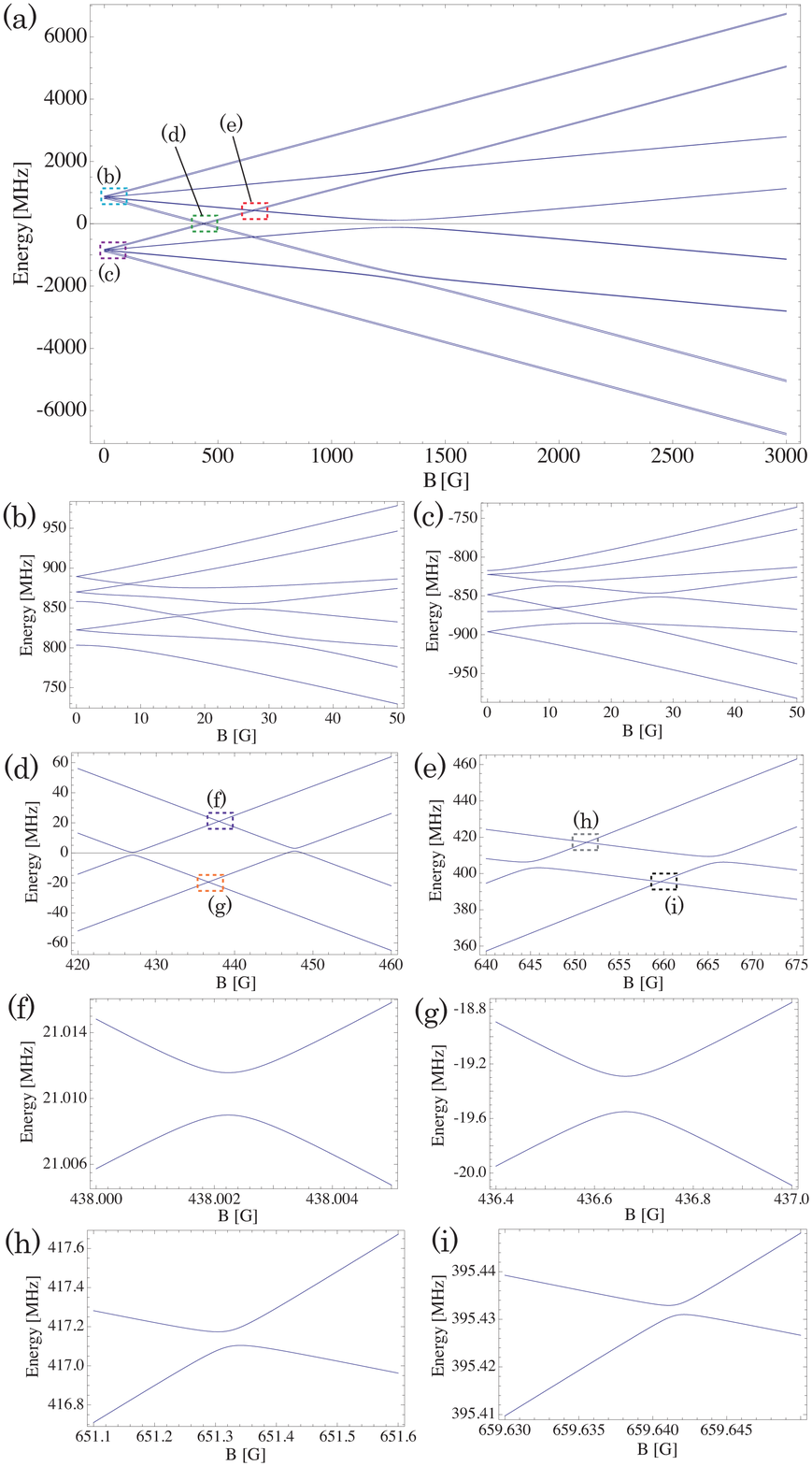}
\end{center}
\vspace{-0.5cm}
\caption{(a) The Zeeman effect for OH in the $v=0$, $X{}^2\Pi_{3/2}$, $J=3/2$ ground state, subject to a bias electric field 
with its strength $E_{\rm DC}=500\,{\rm V/cm}$ and relative angle to the magnetic field $\theta_{BE}=45^{\circ}$. 
(b) The closeups of hyperfine structure for odd parity states (f-states) in zero field and (c) for even parity states (e-states).  
(d) and (e) show the level repulsions between opposite parity states $\ket{{\rm e};M_I=\pm1/2}$ and $\ket{{\rm f};M_I=\pm1/2}$ 
approaching the asymptotic states 
(d) $\ket{{\rm e};M_J=3/2}$ and $\ket{{\rm f};M_J=-3/2}$, and 
(e) $\ket{{\rm e};M_J=3/2}$ and $\ket{{\rm f};M_J=-1/2}$, respectively. 
In contrast to Figures~\ref{FIG_Zeeman3000}(d) and (e), 
all the crossings become avoided crossings, or level repulsions, in (d) and (e) as can be seen from (f)-(i)
due to the field configuration allowing for transitions with $\Delta M_J=0,\, \pm1$. 
}
\label{FIG_Zeeman_S50045}
\end{figure}


\begin{figure}[hptb]
\begin{center}
\includegraphics[width=12cm]{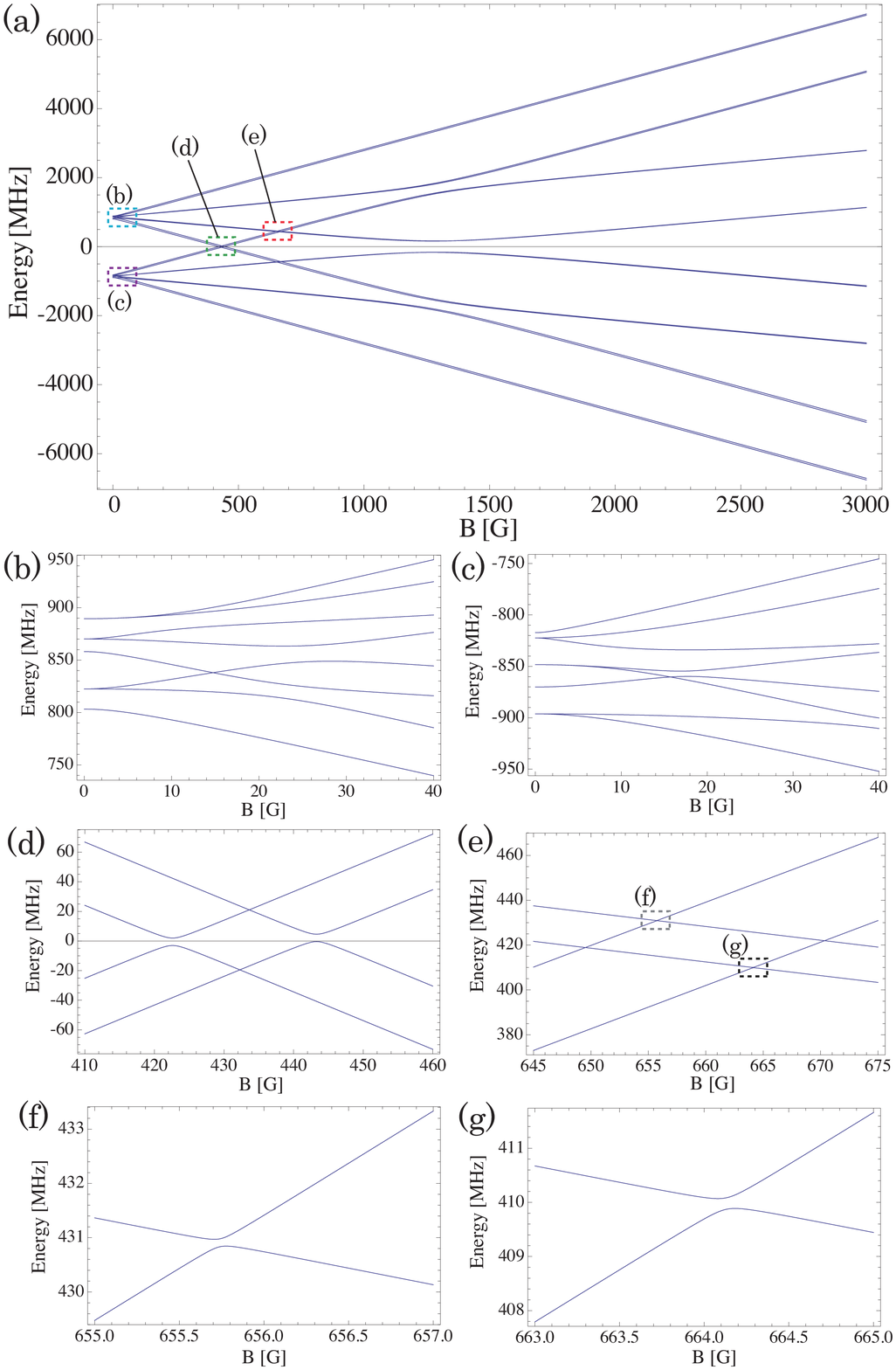}
\end{center}
\vspace{-0.5cm}
\caption{(a) The Zeeman effect for OH in the $v=0$, $X{}^2\Pi_{3/2}$, $J=3/2$ ground state, subject to a bias electric field 
with its strength $E_{\rm DC}=500\,{\rm V/cm}$ and relative angle to the magnetic field $\theta_{BE}=90^{\circ}$. 
(b) The closeups of hyperfine structure for odd parity states (f-states) in zero field and (c) for even parity states (e-states).  
(d) and (e) show the level repulsions between opposite parity states $\ket{{\rm e};M_I=\pm1/2}$ and $\ket{{\rm f};M_I=\pm1/2}$ 
as Figures~\ref{FIG_Zeeman_S50045}(d) and (e), respectively, but 
the absence of a Stark effect with $\Delta M_J=0$ changes 
the upper- and lower-most level repulsions in Figure~\ref{FIG_Zeeman_S50045}(d) 
and the left- and right-most level repulsions in Figure~\ref{FIG_Zeeman_S50045}(e) 
into level crossings. 
There are still level repulsions in (d) and (e) related to transitions with $\Delta M_J=\pm1$, as seen in 
the further closeups (f) and (g). 
}
\label{FIG_Zeeman_S50090}
\end{figure}

We first fix the strength of the electric field $E_{\rm DC}$ and its angle $\theta_{BE}$ 
relative to the magnetic field, as shown in Figure~\ref{FIG_OH_EB}. 
Then, we numerically diagonalize the effective Hamiltonian with external fields Eq.\,(\ref{H_EB}) 
and plot the energy spectrum of the lowest 16 states as a function of the applied magnetic field. 
Figure~\ref{FIG_Zeeman_S00500}(a) shows the spectrum 
with the electric field parallel to the magnetic field, $\theta_{BE}=0^{\circ}$ and strength $E_{\rm DC}=500\,{\rm V/cm}$. 
We observed that Figure~\ref{FIG_Zeeman_S00500}(a) 
is quite similar to the pure Zeeman spectrum Figure~\ref{FIG_Zeeman3000}(a). 
In this case, from the conservation of angular momentum, 
the matrix elements of the Stark term Eq.\,(\ref{HS1}) becomes non-zero only between 
states with $\Delta J=0,\, \pm1$ and $\Delta M_J=0$ (selection rule for $\theta_{BE}=0^{\circ}$). 
On the other hand, states with different parities but same $M_J=0$ are separated by 
the $\Lambda$-doubling splitting $\Delta_{\rm LD}$, 
and the magnetic field keeps them  apart since the Zeeman effect Eq.\,(\ref{par_Z}) does not mix parity states. 
Therefore, one might consider that applying parallel electric and magnetic fields 
does not qualitatively change the pure Zeeman spectrum. 
However, looking at hyperfine structure, 
Figures~\ref{FIG_Zeeman3000}(b)-(g) and Figures~\ref{FIG_Zeeman_S00500}(b)-(g), 
we found that there are qualitative differences. 
In particular, the electric field causes level repulsions 
where both the Stark effects and hyperfine interactions play a significant role. 
First of all, 
the closeups Figures~\ref{FIG_Zeeman_S00500}(b) and (c) differ from 
Figures~\ref{FIG_Zeeman3000}(b) and (c), respectively, 
since the Stark effects break the hyperfine degeneracies in zero fields 
into the degeneracies with $M_F=\pm |M_F|$, as can be seen in Figures~\ref{FIG_Stark3000}(b) and (c). 
Secondly, and even more intriguing, 
there are level repulsions in Figures~\ref{FIG_Zeeman_S00500}(f) and (g), 
which do not appear in the pure Zeeman spectrum Figures~\ref{FIG_Zeeman3000}(f) and (g). 
This is because under the electric field 
the energy eigenstates change from fixed-parity states into fixed-$\Omega$ states
and such fixed-$\Omega$ fractions trigger level repulsions via hyperfine interactions. 
We remark that, due to the conservation of angular momentum, 
the hyperfine interactions have non-zero matrix elements 
between the same parity states with $\Delta M_J=-\Delta M_I$; 
this explains why only one of the four level crossings becomes level repulsion 
in Figures~\ref{FIG_Zeeman_S00500}(f) and (g). 
These level repulsions in the Zeeman spectrum
can be observed only when we take into account both Stark effect and hyperfine interactions, 
thus our appellation Stark-induced hyperfine level repulsions. 
There is no qualitative difference between Figures~\ref{FIG_Zeeman3000}(d), (e) and Figures~\ref{FIG_Zeeman_S00500}(d), (e) 
since the condition for the hyperfine interaction, $\Delta M_J=-\Delta M_I$, is not satisfied. 

In the case of non-parallel electric and magnetic fields, $\theta_{BE}=45^{\circ}$, 
the selection rule of the Stark term Eq.\,(\ref{HS1}) for $\theta_{BE}=45^{\circ}$ 
becomes $\Delta J=0,\, \pm1$ and $\Delta M_J=0,\, \pm1$, 
yielding level repulsions between different parity states with $\Delta M_J=\pm1$ 
around $B\sim1300\,{\rm G}$ of Figure~\ref{FIG_Zeeman_S50045}(a), 
where the strength of the electric field is set to be $E_{\rm DC}=500\,{\rm V/cm}$. 
These level repulsions, directly caused by the Stark effect, 
have energy gaps of 200-$300\,{\rm MHz}$ 
and dominate over Stark-induced hyperfine level repulsions 
whose energy gaps are on the order of $10\,{\rm MHz}$ for $E_{\rm DC}=500\,{\rm V/cm}$. 
There are level repulsions also in Figures~\ref{FIG_Zeeman_S50045}(b) and (c) 
satisfying the selection rule $\Delta M_J=\pm1$, 
that  do not appear in Figures~\ref{FIG_Zeeman_S00500}(b) and (c) for $\theta_{BE}=0^{\circ}$. 
The left- and right-most level repulsions in Figure~\ref{FIG_Zeeman_S50045}(d) 
come from the Stark effect with $\Delta M_J=\pm1$, 
while the left- and right-most level repulsions in Figure~\ref{FIG_Zeeman_S50045}(e) 
from the Stark effect with $\Delta M_J=\pm1$ and $\Delta M_J=0$. 
We found another kind of level repulsion in Figures~\ref{FIG_Zeeman_S50045}(d) and (e), 
see also Figures~\ref{FIG_Zeeman_S50045}(f)-(i). 
The upper- and lower-most level repulsions 
in Figure~\ref{FIG_Zeeman_S50045}~d) require not only 
the Stark effects with $\Delta M_J=\pm1$ and $\Delta M_J=0$
but also the hyperfine interactions; 
and the upper- and lower-most level repulsions in Figure~\ref{FIG_Zeeman_S50045}(e) require 
the Stark effect with $\Delta M_J=\pm1$ and the hyperfine interactions. 
As we will see next, some of these level repulsions become 
level {\it crossings} for $\theta_{BE}=90^{\circ}$ 
since  the Stark effect with $\Delta M_J=0$ vanishes in that case. 

When we apply an electric field perpendicular to the magnetic field, 
the selection rule of the Stark term Eq.\,(\ref{HS1}) is $\Delta J=0,\, \pm1$ and $\Delta M_J=\pm1$. 
Then, the Zeeman spectrum shown in Figure~\ref{FIG_Zeeman_S50090}(a), 
where $\theta_{BE}=90^{\circ}$ and $E_{\rm DC}=500\,{\rm V/cm}$,  
has level repulsions between different parity states with $\Delta M_J=\pm1$ 
around $B\sim1300\,{\rm G}$ as also seen in Figure~\ref{FIG_Zeeman_S50045}(a). 
Also, Figures~\ref{FIG_Zeeman_S50090}(b) and (c) 
are qualitatively the same as Figures~\ref{FIG_Zeeman_S50045}(b) and (c), respectively, 
and their quantitative differences mainly result from differences in the transverse components of the electric field. 
The absence of a Stark effect with $\Delta M_J=0$ changes 
some of the level repulsions into level crossings; 
the upper- and lower-most level repulsions in Figure~\ref{FIG_Zeeman_S50045}(d) 
and the left- and right-most level repulsions in Figure~\ref{FIG_Zeeman_S50045}(e) 
become level crossings, as seen in Figures~\ref{FIG_Zeeman_S50090}(d) and (e).  

\subsection{Comparison with previous work}

\begin{figure}[t]
\begin{center}
\includegraphics[width=16cm]{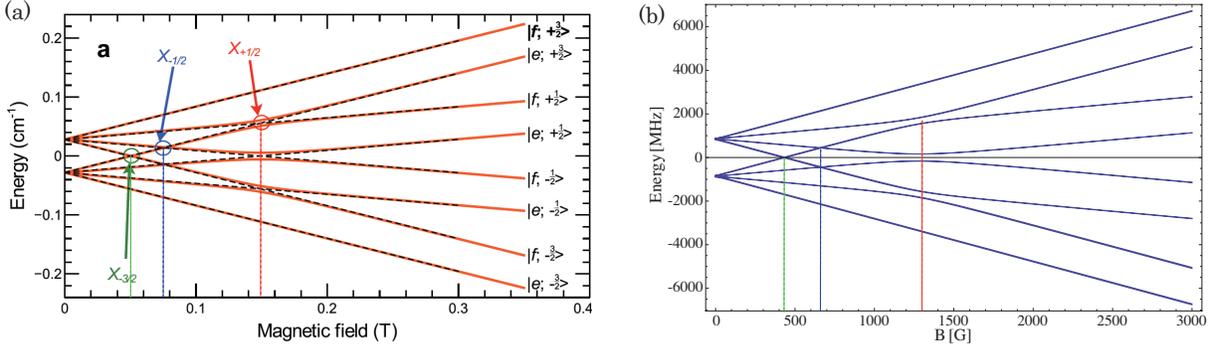}
\end{center}
\vspace{-0.5cm}
\caption{The Zeeman effect or OH in the $v=0$, $X{}^2\Pi_{3/2}$, $J=3/2$ ground state, subject to a bias electric field 
with its strength $E_{\rm DC}=500\,{\rm V/cm}$ and relative angle to the magnetic field $\theta_{BE}=90^{\circ}$. 
(a) is taken from Figure~1 in \cite{Ye_2012_2} 
where $X_i$ labels the crossings of the $\ket{{\rm e};3/2}$ state with the $\ket{{\rm f};M_J=i}$ states ($i=-3/2,-1/2,1/2$), 
and (b) is our calculation, the same as Figure~\ref{FIG_Zeeman_S50090}(a). 
We add vertical dashed lines to compare the values of the magnetic fields at the crossings. 
}
\label{cf_JY}
\end{figure}

Finally, we shall compare our results with previous work on OH molecules. 
Recent works on the single-particle spectrum of OH in the presence of electric and/or magnetic fields 
dealt with phenomenological models which explicitly include the Lambda-doubling splitting $\Delta_{\rm LD}$ 
and restrict themselves only to the lowest 16 states in the ${}^{2}\Pi_{3/2}$ manifold \cite{Bohn_2002,Bohn_2005,Ye_2012_2}, 
or even to  the lowest 8 states neglecting hyperfine structure \cite{Ye_2012_1,Bohn_2013_1,Bohn_2013_2,Bhattacharya_2013}. 
There exists one study which takes into account the effects of the higher rotational states 
and states in the ${}^{2}\Pi_{1/2}$ manifold \cite{Gaerttner_2013}, 
but still it neglects the hyperfine structure, centrifugal distortion effects, 
rotational Zeeman effect, electronic spin anisotropic Zeeman effect, and 
parity-dependent and non-cylindrical Zeeman effects. 
These effects are not negligible when we investigate cold and ultracold physics of OH molecules, 
as shown in Figure~\ref{FIG_energy_scales_H0}. 
As an illustration, we make a comparison of the OH spectrum between the result of \cite{Ye_2012_2} and ours. 
Figure~\ref{cf_JY} shows 
the Zeeman spectrum of OH in the presence of an electric field with strength $E_{\rm DC}=500\,{\rm V/cm}$ 
and relative angle to the magnetic field $\theta_{BE}=90^{\circ}$, 
where Figure~\ref{cf_JY}(a) is taken from \cite{Ye_2012_2} and Figure~\ref{cf_JY}(b) is our result. 
In Figure~\ref{cf_JY}(a), $X_i$ labels the crossings of the $\ket{{\rm e};3/2}$ state 
with the $\ket{{\rm f};M_J=i}$ states ($i=-3/2,-1/2,1/2$). 
We can see that the crossing $X_{-3/2}$ occurs around $B=500\,{\rm G}$ in Figure~\ref{cf_JY}(a), 
while it occurs around $B=430\,{\rm G}$ in Figure~\ref{cf_JY}(b). 
Also, the crossings $X_{-1/2}$ and  $X_{1/2}$ occur around 
$B=750\,{\rm G}$ and $B=1500\,{\rm G}$, respectively in Figure~\ref{cf_JY}(a), 
while they occur around $B=660\,{\rm G}$ and $B=1300\,{\rm G}$, respectively in Figure~\ref{cf_JY}(b). 
The crossing points of $X_i$ are reduced by more than 10 percent in our results 
since the states with $J=3/2$ in the ${}^{2}\Pi_{1/2}$ manifold, 
which are not considered in the reduced model of \cite{Ye_2012_2}, give non-negligible contributions 
via the electronic spin isotropic Zeeman coupling to the states with $J=3/2$ in the ${}^{2}\Pi_{3/2}$ manifold.

\section{Conclusions}

We studied the single-particle energy spectra of the hydroxyl free radical OH 
in its lowest electronic and rovibrational level both in zero field and in combined electric and magnetic fields. 
The hyperfine interactions and centrifugal distortion effects are fully taken into account 
to yield the zero-field spectrum of the lowest ${}^2\Pi_{3/2}$ manifold to an accuracy of less than 2\,kHz; 
in comparison, previous results obtained an accuracy of a few MHz. 
Our more precise calculations are 
necessary to enable accurate investigation of both the single-molecule and many-body physics 
of OH molecules at microKelvin temperatures and below, which we expect will be achieved in the near future. 
We also examined level crossings and repulsions in hyperfine structure 
caused by applied electric and magnetic fields. 
The level repulsions play a significant role in experiments allowing 
for transitions between low-field seeking and high-field seeking states. 
We found that in order to estimate the values of magnetic fields at the level repulsions 
in the ground states of the $J=3/2$, ${}^{2}\Pi_{3/2}$ manifold 
it is necessary to include the coupling with the excited states of the $J=3/2$, ${}^{2}\Pi_{1/2}$ manifold.
In this paper, we dealt only with static electric and magnetic fields, 
leaving the microwave dressing of OH for future work. 
The microwave dressing via AC electric fields makes 
it possible to realize multi-component systems 
with both degeneracy and dipole-dipole interactions between different components, 
and thus explore a rapidly growing field, 
that is, quantum simulation of multi-component dipolar systems with ultracold molecules.




\section*{Acknowledgement}

We thank Yehuda Band, Goulven Qu\'{e}m\'{e}ner, Jun Ye for helpful discussions.  
This research was supported in part by AFOSR Grant No.~FA9550-11-1-0224 
and by the NSF under Grants PHY-1207881 and NSF PHY11-25915. 
We appreciate the Aspen Center for Physics, supported in part by the NSF Grant No.~1066293, 
for hospitality during the writing of this paper.


\appendix

\section{Spherical tensor operators and related formulae}
\label{sec:stp}

Let $\hat{\bf A}$, $\hat{\bf B}$ be arbitrary vector operators with their space-fixed components 
$(\hat{A}_X,\hat{A}_Y,\hat{A}_Z)$ and $(\hat{B}_X,\hat{B}_Y,\hat{B}_Z)$. 
Note that we use indices $X$, $Y$, $Z$ for space-fixed components 
and $x$, $y$, $z$ for molecule-fixed components of vector operators. 
We define 
ladder operators as usual, $\hat{A}_{\pm}=\hat{A}_X\pm{\rm i}\hat{A}_Y$. 
Then the rank-1 (irreducible) spherical tensor operator $T^1(\hat{\bf A})$ is given by 
\beq
T^1_0(\hat{\bf A})
&=&
\hat{A}_Z
~,\qquad
T^1_1(\hat{\bf A})
~=~
-\frac{1}{\sqrt{2}}\hat{A}_{+}
~,\qquad
T^1_{-1}(\hat{\bf A})
~=~
\frac{1}{\sqrt{2}}\hat{A}_{-}
~.
\eeq
The scalar product of two rank-1 spherical tensor operators $T^{1}(\hat{\bf A})$ and $T^{1}(\hat{\bf B})$ becomes 
\beq
T^1(\hat{\bf A})\cdot T^1(\hat{\bf B})
&=&
\sum_{p=0,\pm 1}(-1)^p
T^1_p(\hat{\bf A})T^1_{-p}(\hat{\bf B})
~=~
\sum_{q=0,\pm 1}(-1)^q
T^1_q(\hat{\bf A})T^1_{-q}(\hat{\bf B})
~, 
\eeq
where we specified space-fixed and molecule-fixed components by $p$ and $q$, respectively. 
Also, the irreducible tensor product of 
two rank-1 spherical tensor operators $T^{1}(\hat{\bf A})$ and $T^{1}(\hat{\bf B})$ is given by
\beq
T_0^{2}(\hat{\bf A},\hat{\bf B})
&=&
\frac{1}{\sqrt{6}}
\left[
T^1_1(\hat{\bf A})T^1_{-1}(\hat{\bf B})
+
2T^1_0(\hat{\bf A})T^1_{0}(\hat{\bf B})
+
T^1_{-1}(\hat{\bf A})T^1_{1}(\hat{\bf B})
\right]
~,\\
T_{\pm1}^{2}(\hat{\bf A},\hat{\bf B})
&=&
\frac{1}{\sqrt{2}}
\left[
T^1_{\pm 1}(\hat{\bf A})T^1_{0}(\hat{\bf B})
+
T^1_{0}(\hat{\bf A})T^1_{\pm 1}(\hat{\bf B})
\right]
~,\\
T_{\pm 2}^{2}(\hat{\bf A},\hat{\bf B})
&=&
T^1_{\pm 1}(\hat{\bf A})T^1_{\pm 1}(\hat{\bf B})
~.
\eeq 
For a general definition of spherical tensor operators and their tensor products, 
see \cite{BrownCarrington_2003,Zare_1988}. 
A spherical tensor operator in the space-fixed frame, $T^{1}_p(\hat{\bf A})$, is 
related to its representation in the molecule-fixed frame, $T^{1}_q(\hat{\bf A})$, via 
the matrix elements of the Wigner $D$-matrix $\mathscr{D}^{(1)}_{p,q}$, 
\beq
T^1_q(\hat{\bf A})
&=&
\sum_{p=0,\pm 1}
 \mathscr{D}_{p,q}^{(1)}(\omega)
T^1_p(\hat{\bf A})
~, 
\eeq
or equivalently, 
\beq
T^1_p(\hat{\bf A})
&=&
\sum_{q=0,\pm 1}
 \mathscr{D}_{p,q}^{(1)\ast}(\omega)
T^1_q(\hat{\bf A})
~.
\eeq 
Here Euler angles $\omega=(\phi,\theta,\chi)$ define a general orientation of the molecule-fixed frame. 
In the following we will use the Wigner $3j$-symbol, defined as 
\beq
\biggl(\begin{array}{ccc}
j_1 & j_2 & j_3 \\
m_1 & m_2& m_3
\end{array}\biggl)
&=&
(-1)^{j_1-j_2-m_3}(2j_3+1)^{-1/2}\braket{j_1,m_1,j_2,m_2|j_3,-m_3}
~,
\eeq
where $\braket{j_1,m_1,j_2,m_2|j_3,-m_3}$ is a Clebsch-Gordan coefficient. 
The Wigner-Eckart theorem gives the matrix elements of the spherical tensor operators, 
\beq\fl
\bra{J',\Omega',M'_{J'}}T^1_p(\hat{\bf  J})\ket{J,\Omega, M_J}
&=&
\delta_{J,J'}\delta_{\Omega,\Omega'}
(-1)^{J-M'_{J'}}\sqrt{J(J+1)(2J+1)}
\biggl(\begin{array}{ccc}
J & 1 & J \\
-M'_{J'} & p & M_J
\end{array}\biggl)
~, \qquad
\eeq
and 
\beq\fl
\lefteqn{
\bra{J',\Omega', M'_{J'}}
T^2_{p}(\hat{\bf  J},\hat{\bf  J})
\ket{J,\Omega, M_J}
}
\nonumber\\
\hspace{-70pt}
=~
\delta_{J,J'}\delta_{M_J,M'_{J'}}
(-1)^{J-\Omega'}
\frac{1}{2\sqrt{6}}
\sqrt{(2J-1)2J(2J+1)(2J+2)(2J+3)}
\biggl(\begin{array}{ccc}
J & 2 & J \\
-\Omega' & p & \Omega
\end{array}\biggl)
~,
\eeq
in the space-fixed frame, and 
\beq\fl
\bra{J',\Omega',M'_{J'}}T^1_q(\hat{\bf  J})\ket{J,\Omega, M_J}
&=&
\delta_{J,J'}\delta_{M_J,M'_{J,}}
(-1)^{J-\Omega'}\sqrt{J(J+1)(2J+1)}
\biggl(\begin{array}{ccc}
J & 1 & J \\
-\Omega' & q & \Omega
\end{array}\biggl)
~,
\eeq
in the molecule-fixed frame, 
where $\ket{J,\Omega, M_J}$ 
is the simultaneous eigenstate of operators $\hat{\bf  J}^2$, $\hat{J}_z$, and $\hat{J}_Z$. 
Then, it is straightforward to have matrix elements of the nuclear spin operator and electronic spin operator, 
\beq
\bra{I,M'_{I}}
T^1_{p}(\hat{\bf I})
\ket{I,M_I}
&=&
(-1)^{I-M'_{I}}\sqrt{I(I+1)(2I+1)}
\biggl(\begin{array}{ccc}
I & 1 & I \\
-M'_{I} & p & M_I
\end{array}\biggl)~, 
\eeq
and
\beq
 \bra{S,\Sigma'}T^1_q(\hat{\bf S})\ket{S,\Sigma}
 &=&
(-1)^{S-\Sigma'}\sqrt{S(S+1)(2S+1)}
\biggl(\begin{array}{ccc}
S & 1 & S \\
-\Sigma' & q& \Sigma
\end{array}\biggl)
~, 
\eeq
where $\ket{I,M_I}$ and $\ket{S,\Sigma}$ 
are the simultaneous eigenstates of operators $\hat{\bf I}^2$, $\hat{I}_Z$ and $\hat{\bf S}^2$, $\hat{S}_z$, respectively. 
We need a few more formulae to calculate the matrix elements of the effective Hamiltonian for OH molecule, 
\beq
\lefteqn{
\bra{J',\Omega',M'_{J'}}
 \mathscr{D}_{p,q}^{(1)\ast}(\omega)
\ket{J,\Omega ,M_J}
}
\nonumber\\
&=&
(-1)^{M'_{J'}-\Omega'}\sqrt{(2J'+1)(2J+1)}
\biggl(\begin{array}{ccc}
J' & 1 & J \\
-\Omega' & q & \Omega
\end{array}\biggl)
\biggl(\begin{array}{ccc}
J' & 1 & J \\
-M'_{J'} & p & M_J
\end{array}\biggl)
~, 
\eeq
and
\beq
\lefteqn{
\bra{J',\Omega',M'_{J'}}
\sum_{p=0,\pm 1}(-1)^p
T^1_p(\hat{\bf  J})
 \mathscr{D}_{-p,q}^{(1)\ast}(\omega)
 \ket{J,\Omega, M_J}
}
\nonumber\\
&=&
\delta_{J,J'}\delta_{M_J,M'_{J'}}
(-1)^{J-\Omega'}\sqrt{J(J+1)(2J+1)}
\biggl(\begin{array}{ccc}
J & 1 & J \\
-\Omega' & q& \Omega
\end{array}\biggl)
~.
\eeq
We finally remark that due to the anomalous commutation relation in the molecule-fixed frame, 
\beq
[\hat{J}_{\alpha},\hat{J}_{\beta}]
&=&-i\epsilon_{\alpha\beta\gamma}\hat{J}_{\gamma}
\qquad(\alpha,\beta,\gamma\in\{x,y,z\})
~, 
\eeq
we should replace the the molecule-fixed components of spherical tensor operators as
$T^{1}_q(\hat{\bf  J})\to(-1)^qT^{1}_{-q}\,(\hat{\bf  J})$ before calculating matrix elements. 
A detailed derivation of the above formulae may be found in \cite{BrownCarrington_2003,Zare_1988}.

\section{Matrix elements of the effective Hamiltonian}
\label{sec:matele}


We have chosen a Hund's case (a) basis in the molecule-fixed frame 
to obtain a representation of the molecular Hamiltonian,  
\beq
\ket{L\Lambda,\! S\Sigma, \! J\Omega M_J, \! IM_I}
~=~
\ket{L,\Lambda}\ket{S,\Sigma}\ket{J,\Omega, M_J}\ket{I,M_I}
\eeq
with quantum numbers 
$L=1$, $\Lambda=\pm1$, $S=1/2$, $\Sigma=\pm1/2$, 
$J\geq1/2$, $\Omega=\Lambda+\Sigma$, $-J\leq M_J\leq J$, $I=1/2$, and $M_I=\pm1/2$. 
Based on the formulae given in \ref{sec:stp}, we can derive the following 23 matrix elements 
of the effective Hamiltonian for OH. 

(1) Spin-orbit coupling
\beq \fl
\lefteqn{
\bra{L\Lambda' \!,\! S\Sigma' \!,\! J'\Omega' M'_{J'} ,\! IM'_I}
A_{\rm so}  T^1_{q=0}(\hat{\bf L})T^1_{q=0}(\hat{\bf S})
\ket{L\Lambda,\! S\Sigma, \! J\Omega M_J, \! IM_I}}
\nonumber\\
\hspace{-70pt}
=~
A_{\rm so} \Lambda\Sigma
\delta_{\Lambda,\Lambda'}\delta_{\Sigma,\Sigma'}
\delta_{J,J'}\delta_{\Omega,\Omega'}\delta_{M_J,M'_{J'}}
\delta_{M_I,M'_{I'}}
\, .
\eeq

(2) Molecular rotation
\beq\fl
\lefteqn{
\bra{L\Lambda' \!,\! S\Sigma' \!,\! J'\Omega' M'_{J'} ,\! IM'_I}
B_{N}\hat{\bf N}^2
\ket{L\Lambda,\! S\Sigma, \! J\Omega M_J, \! IM_I}
}
\nonumber\\
\hspace{-70pt}
=~
B_{N}\left[
J(J+1)+S(S+1)-2\Omega\Sigma
\right]
\delta_{\Lambda,\Lambda'}\delta_{\Sigma,\Sigma'}
\delta_{J,J'}\delta_{\Omega,\Omega'}\delta_{M_J,M'_{J'}}
\delta_{M_I,M'_{I'}}
\nonumber\\
\hspace{-70pt}~
-2B_{N}
\delta_{\Lambda,\Lambda'}
\delta_{J,J'}\delta_{M_J,M'_{J'}}
\delta_{M_I,M'_{I'}}
\nonumber\\
\hspace{-70pt}\quad
\times
(-1)^{J-\Omega'+S-\Sigma'}
\sqrt{J(J+1)(2J+1)S(S+1)(2S+1)}
\sum_{q=\pm 1}
\biggl(\begin{array}{ccc}
J & 1 & J \\
-\Omega' & q& \Omega
\end{array}\biggl)
\biggl(\begin{array}{ccc}
S & 1 & S \\
-\Sigma' & q& \Sigma
\end{array}\biggl)
\, .
\eeq

(3) Spin-molecular rotation coupling 
\beq\fl
\lefteqn{
\bra{L\Lambda' \!,\! S\Sigma' \!,\! J'\Omega' M'_{J'} ,\! IM'_I}
\gamma\, T^1(\hat{\bf  J}-\hat{\bf S})\cdot T^1(\hat{\bf S})
\ket{L\Lambda,\! S\Sigma, \! J\Omega M_J, \! IM_I}
}\nonumber\\
\hspace{-70pt}
=~
\gamma\left[
\Omega\Sigma- S(S+1)
\right]
\delta_{\Lambda,\Lambda'}\delta_{\Sigma,\Sigma'}
\delta_{J,J'}\delta_{\Omega,\Omega'}\delta_{M_J,M'_{J'}}
\delta_{M_I,M'_{I'}}
\nonumber\\
\hspace{-70pt}~
+\gamma
\delta_{\Lambda,\Lambda'}
\delta_{J,J'}\delta_{M_J,M'_{J'}}
\delta_{M_I,M'_{I'}}
\nonumber\\
\hspace{-70pt}\quad
\times
(-1)^{J-\Omega'+S-\Sigma'}
\sqrt{J(J+1)(2J+1)S(S+1)(2S+1)}
\sum_{q=\pm 1}
\biggl(\begin{array}{ccc}
J & 1 & J \\
-\Omega' & q & \Omega
\end{array}\biggl)
\biggl(\begin{array}{ccc}
S & 1 & S \\
-\Sigma' & q& \Sigma
\end{array}\biggl)
\, .
\eeq

(4) $\Lambda$-doubling term 
\beq
\fl
\lefteqn{
\bra{L\Lambda' \!,\! S\Sigma' \!,\! J'\Omega' M'_{J'} ,\! IM'_I}
\sum_{q=\pm 1}{\rm e}^{-2{\rm i}q\phi}\,
\left[
-Q\, T^2_{2q}(\hat{\bf  J},\hat{\bf  J})+(P+2Q)T^2_{2q}(\hat{\bf  J},\hat{\bf S})
\right]
\ket{L\Lambda,\! S\Sigma, \! J\Omega M_J, \! IM_I}
}\nonumber\\
\hspace{-70pt}
=~
\delta_{J,J'}\delta_{M_{J},M'_{J'}}\delta_{M_{I},M'_{I}}
(-1)^{J-\Omega'}\sqrt{J(2J+1)}
\sum_{q=\pm 1}
\delta_{\Lambda',\Lambda-2q}
\nonumber\\
\hspace{-70pt}
~\times
\biggl\{
\frac{Q}{2\sqrt{3}}
\delta_{\Sigma,\Sigma'}
\sqrt{(2J-1)(2J+2)(2J+3)}
\biggl(\begin{array}{ccc}
J & 2 & J \\
-\Omega' & -2q& \Omega
\end{array}\biggl)
\nonumber\\
\hspace{-70pt}\qquad
+(P+2Q)
(-1)^{S-\Sigma'}
\sqrt{(J+1)S(S+1)(2S+1)}
\biggl(\begin{array}{ccc}
J & 1 & J \\
-\Omega' & -q& \Omega
\end{array}\biggl)
\biggl(\begin{array}{ccc}
S & 1 & S \\
-\Sigma' & q& \Sigma
\end{array}\biggl)
\biggl\}
\, .
\eeq

(5) Magnetic hyperfine interaction (I)
\beq\fl
\lefteqn{
\bra{L\Lambda' \!,\! S\Sigma' \!,\! J'\Omega' M'_{J'} ,\! IM'_I}
a\,T^1_{q=0}(\hat{\bf I})T^1_{q=0}(\hat{\bf L})
\ket{L\Lambda,\! S\Sigma, \! J\Omega M_J, \! IM_I}
}\nonumber\\
\hspace{-70pt}
=~
a\Lambda
\delta_{\Lambda,\Lambda'}
\delta_{\Sigma,\Sigma'}
(-1)^{M'_{J'}-\Omega'+I-M'_{I}}
\sqrt{I(I+1)(2I+1)(2J'+1)(2J+1)}
\biggl(\begin{array}{ccc}
J' & 1 & J \\
-\Omega' & 0 & \Omega
\end{array}\biggl)
\nonumber\\
\hspace{-70pt}
~\times
\sum_{p=0,\pm 1}
(-1)^p
\biggl(\begin{array}{ccc}
J' & 1 & J \\
-M'_{J'} & p & M_J
\end{array}\biggl)
\biggl(\begin{array}{ccc}
I & 1 & I \\
-M'_{I} & -p & M_I
\end{array}\biggl)
\, . 
\eeq

(6) Magnetic hyperfine interaction (II)
\beq\fl
\lefteqn{
\bra{L\Lambda' \!,\! S\Sigma' \!,\! J'\Omega' M'_{J'} ,\! IM'_I}
b_F T^1(\hat{\bf I})\cdot T^1(\hat{\bf S})
\ket{L\Lambda,\! S\Sigma, \! J\Omega M_J, \! IM_I}
}\nonumber\\
\hspace{-70pt}
=~
b_F
\delta_{\Lambda,\Lambda'}
(-1)^{M'_{J'}-\Omega'+I-M'_{I}+S-\Sigma'}
\sqrt{(2J'+1)(2J+1)I(I+1)(2I+1)S(S+1)(2S+1)}
\nonumber\\
\hspace{-70pt}
~\times
\sum_{p,q=0,\pm1} (-1)^p
\biggl(\begin{array}{ccc}
J' & 1 & J \\
-\Omega' & q & \Omega
\end{array}\biggl)
\biggl(\begin{array}{ccc}
J' & 1 & J \\
-M'_{J'} & p & M_J
\end{array}\biggl)
\biggl(\begin{array}{ccc}
I & 1 & I \\
-M'_{I} & -p & M_I
\end{array}\biggl)
\biggl(\begin{array}{ccc}
S & 1 & S \\
-\Sigma' & q & \Sigma
\end{array}\biggl)
\, .
\eeq

(7) Magnetic hyperfine interaction (III)
\beq\fl
\lefteqn{
\bra{L\Lambda' \!,\! S\Sigma' \!,\! J'\Omega' M'_{J'} ,\! IM'_I}
\biggl\{
\sqrt{\frac{2}{3}}c\, T^2_{q=0}(\hat{\bf I},\hat{\bf S})
+d\sum_{q=\pm 1} {\rm e}^{-2{\rm i}q\phi}\, T^2_{2q}(\hat{\bf I},\hat{\bf S})
\biggl\}
\ket{L\Lambda,\! S\Sigma, \! J\Omega M_J, \! IM_I}
}\nonumber\\
\hspace{-70pt}
=~
(-1)^{M'_{J'}-\Omega' +I-M'_{I} +S-\Sigma'}
\sqrt{(2J'+1)(2J+1)I(I+1)(2I+1)S(S+1)(2S+1)}
\nonumber\\
\hspace{-70pt}
~\times
\sum_{p=0,\pm1}(-1)^{p}
\biggl(\begin{array}{ccc}
J' & 1 & J \\
 -M'_{J'} & p & M_J
\end{array}\biggl)
\biggl(\begin{array}{ccc}
I & 1 & I \\
 -M'_{I} & -p & M_I
\end{array}\biggl)
\nonumber\\
\hspace{-70pt}
~\times
\biggl\{
\sqrt{\frac{10}{3}}c\,  \delta_{\Lambda,\Lambda'}
\sum_{q_1=0,\pm1}(-1)^{q_1}
\biggl(\begin{array}{ccc}
 1& 2 & 1 \\
 -q_1 & 0 & q_1
\end{array}\biggl)
\biggl(\begin{array}{ccc}
J' & 1 & J \\
 -\Omega' & q_1 & \Omega
\end{array}\biggl)
\biggl(\begin{array}{ccc}
S & 1 & S \\
 -\Sigma' & q_1 & \Sigma
\end{array}\biggl)
\nonumber\\
\hspace{-70pt}
~\qquad
+d
\sum_{q=\pm 1}
\delta_{\Lambda',\Lambda-2q} 
\biggl(\begin{array}{ccc}
J' & 1 & J \\
 -\Omega' & -q & \Omega
\end{array}\biggl)
\biggl(\begin{array}{ccc}
S & 1 & S \\
 -\Sigma' & q & \Sigma
\end{array}\biggl)
\biggl\}
\, .
\eeq

(8) Magnetic hyperfine interaction (IV)
\beq\fl
\lefteqn{
\bra{L\Lambda' \!,\! S\Sigma' \!,\! J'\Omega' M'_{J'} ,\! IM'_I}
c_I T^1(\hat{\bf I})\cdot T^1(\hat{\bf  J}-\hat{\bf S})
\ket{L\Lambda,\! S\Sigma, \! J\Omega M_J, \! IM_I}
}\\
\hspace{-70pt}
=~
c_I
\delta_{\Lambda,\Lambda'}
(-1)^{I-M'_{I}}
\sqrt{I(I+1)(2I+1)}
\sum_{p=0,\pm1}(-1)^{p}
\biggl(\begin{array}{ccc}
J' & 1 & J \\
-M'_{J'} & p & M_J
\end{array}\biggl)
\biggl(\begin{array}{ccc}
I & 1 & I \\
-M'_{I} & -p & M_I
\end{array}\biggl)
\nonumber\\
\hspace{-70pt}~
\times
\biggl\{
\delta_{\Sigma,\Sigma'}
\delta_{J,J'}
\delta_{\Omega,\Omega'}
(-1)^{J-M'_{J'}}
\sqrt{J(J+1)(2J+1)}
\nonumber\\
\hspace{-70pt}\qquad
-
(-1)^{M'_{J'}-\Omega'+S-\Sigma'}
\sqrt{(2J'+1)(2J+1)S(S+1)(2S+1)}
\sum_{q=0,\pm1}
\biggl(\begin{array}{ccc}
J' & 1 & J \\
-\Omega' & q & \Omega
\end{array}\biggl)
\biggl(\begin{array}{ccc}
S & 1 & S \\
-\Sigma' & q & \Sigma
\end{array}\biggl)
\biggl\}
\, .
\nonumber
\eeq

(9) Magnetic hyperfine interaction (V)
\beq\fl
\lefteqn{
\bra{L\Lambda' \!,\! S\Sigma' \!,\! J'\Omega' M'_{J'} ,\! IM'_I}
c_I' 
\sum_{q=\pm 1} {\rm e}^{-2{\rm i}q\phi}\,\frac{1}{2}
\left[
T^2_{2q}(\hat{\bf I},\hat{\bf  J}-\hat{\bf S})+T^2_{2q}(\hat{\bf  J}-\hat{\bf S},\hat{\bf I})
\right]
\ket{L\Lambda,\! S\Sigma, \! J\Omega M_J, \! IM_I}
}\\
\hspace{-70pt}
=
-\frac{c_I' }{2}
(-1)^{I-M'_{I}}
\sqrt{I(I+1)(2I+1)(2J'+1)(2J+1)}
\sum_{p=0,\pm1}(-1)^{p}
\biggl(\begin{array}{ccc}
I & 1 & I \\
-M'_{I} & -p & M_I
\end{array}\biggl)
\biggl(\begin{array}{ccc}
J' & 1 & J \\
-M'_{J'} & p & M_J
\end{array}\biggl)
\nonumber\\
\hspace{-70pt}
~\times
\sum_{q=\pm 1}
\delta_{\Lambda',\Lambda-2q} 
\biggl\{
2(-1)^{S-\Sigma'+M'_{J'}-\Omega'}
\sqrt{S(S+1)(2S+1)}
\biggl(\begin{array}{ccc}
S & 1 & S \\
-\Sigma' & q & \Sigma
\end{array}\biggl)
\biggl(\begin{array}{ccc}
J' & 1 & J \\
-\Omega' & -q & \Omega
\end{array}\biggl)
\nonumber\\
\hspace{-70pt}
~+
\delta_{\Sigma,\Sigma'}
\theta(3/2-|\Omega'+q|)
(-1)^{J+M'_{J'}}
\sqrt{J(J+1)(2J+1)}
\biggl(\begin{array}{ccc}
J' & 1 & J \\
-\Omega' & -q & \Omega'+q
\end{array}\biggl)
\biggl(\begin{array}{ccc}
J & 1 & J \\
-\Omega'-q & -q & \Omega
\end{array}\biggl)
\nonumber\\
\hspace{-70pt}
~+
\delta_{\Sigma,\Sigma'}
\theta(3/2-|\Omega'+q|)
(-1)^{J'+M'_{J'}}
\sqrt{J'(J'+1)(2J'+1)}
\biggl(\begin{array}{ccc}
J' & 1 & J' \\
-\Omega' & -q & \Omega'+q
\end{array}\biggl)
\biggl(\begin{array}{ccc}
J' & 1 & J \\
-\Omega'-q & -q & \Omega
\end{array}\biggl)
\biggl\}
\nonumber
\, .
\eeq

(10) Centrifugal distortion effect to molecular rotation (I)
\beq\fl
\lefteqn{
-\bra{L\Lambda' \!,\! S\Sigma' \!,\! J'\Omega' M'_{J'} ,\! IM'_I}
D(\hat{\bf N}^2)^2
\ket{L\Lambda,\! S\Sigma, \! J\Omega M_J, \! IM_I}
}\nonumber\\
\hspace{-70pt}
=
-D\delta_{\Lambda,\Lambda'}\delta_{J,J'}\delta_{M_J,M'_{J'}}\delta_{M_I,M'_{I}}
\nonumber\\
\hspace{-70pt}~
\times
\biggl\{
\delta_{\Omega,\Omega'}\delta_{\Sigma,\Sigma'}
\left[
J(J+1)+S(S+1)-2\Omega\Sigma
\right]^2
\nonumber\\
\hspace{-70pt}\qquad
-2
(-1)^{J-\Omega'+S-\Sigma'}
\sqrt{J(J+1)(2J+1)S(S+1)(2S+1)}
\nonumber\\
\hspace{-70pt}~\qquad
\times
\left[
2J(J+1)+2S(S+1)-2\Omega'\Sigma'-2\Omega\Sigma
\right]
\sum_{q=\pm 1}
\biggl(\begin{array}{ccc}
J & 1 & J \\
-\Omega' & q& \Omega
\end{array}\biggl)
\biggl(\begin{array}{ccc}
S & 1 & S \\
-\Sigma' & q& \Sigma
\end{array}\biggl)
\nonumber\\
\hspace{-70pt}\qquad
+4
\delta_{\Omega,\Omega'}\delta_{\Sigma,\Sigma'}
J(J+1)(2J+1)S(S+1)(2S+1)
\nonumber\\
\hspace{-70pt}\qquad~
\nonumber\\
\hspace{-70pt}\qquad
\times
\sum_{q=\pm 1}
\biggl[
\biggl(\begin{array}{ccc}
J & 1 & J \\
-\Omega & q& \Omega-q
\end{array}\biggl)
\biggl(\begin{array}{ccc}
S & 1 & S \\
-\Sigma & q& \Sigma-q
\end{array}\biggl)
\biggl]^2
\theta(3/2-|\Omega-q|)
\biggl\}
\, .
\eeq

(11) Centrifugal distortion effect to molecular rotation (II)
\beq\fl
\lefteqn{
\bra{L\Lambda' \!,\! S\Sigma' \!,\! J'\Omega' M'_{J'} ,\! IM'_I}
H(\hat{\bf N}^2)^3
\ket{L\Lambda,\! S\Sigma, \! J\Omega M_J, \! IM_I}
}\nonumber\\
\hspace{-70pt}
=~
H\delta_{\Lambda,\Lambda'}\delta_{J,J'}\delta_{M_J,M'_{J'}}\delta_{M_I,M'_{I}}
\nonumber\\
\hspace{-70pt}
\times
\biggl\{
\delta_{\Omega,\Omega'}\delta_{\Sigma,\Sigma'}
\left[
J(J+1)+S(S+1)-2\Omega\Sigma
\right]^3
\nonumber\\
\hspace{-70pt}
~\quad
+4
\delta_{\Omega,\Omega'}\delta_{\Sigma,\Sigma'}
J(J+1)(2J+1)S(S+1)(2S+1)
\sum_{q=\pm 1}\theta(3/2-|\Omega-q|)
\nonumber\\
\hspace{-70pt}
~\qquad\times
\left[
3J(J+1)+3S(S+1)-4\Omega\Sigma-2(\Omega-q)(\Sigma-q)
\right]
\biggl[
\biggl(\begin{array}{ccc}
J & 1 & J \\
-\Omega & q& \Omega-q
\end{array}\biggl)
\biggl(\begin{array}{ccc}
S & 1 & S \\
-\Sigma & q& \Sigma-q
\end{array}\biggl)
\biggl]^2
\nonumber\\
\hspace{-70pt}
~\quad
-2
(-1)^{J-\Omega'+S-\Sigma'}
\sqrt{J(J+1)(2J+1)S(S+1)(2S+1)}
\sum_{q=\pm 1}
\biggl(\begin{array}{ccc}
J & 1 & J \\
-\Omega' & q& \Omega
\end{array}\biggl)
\biggl(\begin{array}{ccc}
S & 1 & S \\
-\Sigma' & q& \Sigma
\end{array}\biggl)
\nonumber\\
\hspace{-70pt}
~\qquad\times
\biggl[
\left[
J(J+1)+S(S+1)-2\Omega'\Sigma'
\right]^2
+
\left[
J(J+1)+S(S+1)-2\Omega\Sigma
\right]^2
\nonumber\\
\hspace{-70pt}
~\qquad\qquad+
\left[
J(J+1)+S(S+1)-2\Omega'\Sigma'
\right]
\left[
J(J+1)+S(S+1)-2\Omega\Sigma
\right]
\nonumber\\
\hspace{-70pt}
~\qquad\qquad+
4J(J+1)(2J+1)S(S+1)(2S+1)
\biggl(\begin{array}{ccc}
J & 1 & J \\
-\Omega' & q& \Omega
\end{array}\biggl)^2
\biggl(\begin{array}{ccc}
S & 1 & S \\
-\Sigma' & q& \Sigma
\end{array}\biggl)^2
\biggl]
\biggl\}
\, .
\eeq

(12) Centrifugal distortion effect to spin-molecular rotation coupling
\beq\fl
\lefteqn{
\bra{L\Lambda' \!,\! S\Sigma' \!,\! J'\Omega' M'_{J'} ,\! IM'_I}
\gamma_D\, \left\{T^1(\hat{\bf  J}-\hat{\bf S})\cdot T^1(\hat{\bf S})\right\}\hat{\bf N}^2
\ket{L\Lambda,\! S\Sigma, \! J\Omega M_J, \! IM_I}
}\nonumber\\
\hspace{-70pt}
=~
\gamma_D
\delta_{\Lambda,\Lambda'}\delta_{J,J'}\delta_{M_J,M'_{J'}}\delta_{M_I,M'_{I}}
\nonumber\\
\hspace{-70pt}~
\times
\biggl\{
\delta_{\Omega,\Omega'}\delta_{\Sigma,\Sigma'}
\left[
\Omega\Sigma- S(S+1)
\right]
\left[
J(J+1)+S(S+1)-2\Omega\Sigma
\right]
\nonumber\\
\hspace{-70pt}
\qquad+
\left[
J(J+1)+3S(S+1)-2\Omega'\Sigma'-2\Omega\Sigma
\right]
\nonumber\\
\hspace{-70pt}~\qquad
\times
(-1)^{J-\Omega'+S-\Sigma'}
\sqrt{J(J+1)(2J+1)S(S+1)(2S+1)}
\sum_{q=\pm 1}
\biggl(\begin{array}{ccc}
J & 1 & J \\
-\Omega' & q& \Omega
\end{array}\biggl)
\biggl(\begin{array}{ccc}
S & 1 & S \\
-\Sigma' & q& \Sigma
\end{array}\biggl)
\nonumber\\
\hspace{-70pt}
\qquad
-2
\delta_{\Omega,\Omega'}\delta_{\Sigma,\Sigma'}
J(J+1)(2J+1)S(S+1)(2S+1)
\nonumber\\
\hspace{-70pt}
~\qquad\times
\sum_{q=\pm 1}
\theta(3/2-|\Omega-q|)
\biggl[
\biggl(\begin{array}{ccc}
J & 1 & J \\
-\Omega & q& \Omega-q
\end{array}\biggl)
\biggl(\begin{array}{ccc}
S & 1 & S \\
-\Sigma & q& \Sigma-q
\end{array}\biggl)
\biggl]^2
\biggl\}
\, .
\eeq

(13) Centrifugal distortion effect to $\Lambda$-doubling term (I) 

\textbullet~ $-Q_D$ term
\beq\fl
\lefteqn{
-\bra{L\Lambda' \!,\! S\Sigma' \!,\! J'\Omega' M'_{J'} ,\! IM'_I}
\sum_{q=\pm 1}{\rm e}^{-2{\rm i}q\phi}\,
\frac{Q_D}{2}
\left[
T^2_{2q}(\hat{\bf  J},\hat{\bf  J})\hat{\bf N}^2+\hat{\bf N}^2T^2_{2q}(\hat{\bf  J},\hat{\bf  J})
\right]
\ket{L\Lambda,\! S\Sigma, \! J\Omega M_J, \! IM_I}
}\nonumber\\
\hspace{-70pt}
=~
\frac{Q_D}{4\sqrt{6}}
\delta_{J,J'}\delta_{M_J,M'_{J'}}\delta_{M_I,M'_{I}}
\sqrt{(2J-1)2J(2J+1)(2J+2)(2J+3)}
\sum_{q=\pm 1}
\delta_{\Lambda',\Lambda-2q}
\nonumber\\
\hspace{-70pt}
\times
\biggl\{
\delta_{\Sigma,\Sigma'}
\left[
2J(J+1)+2S(S+1)-2(\Omega'+\Omega)\Sigma
\right]
(-1)^{J-\Omega'}
\biggl(\begin{array}{ccc}
J & 2 & J \\
-\Omega' & -2q& \Omega
\end{array}\biggl)
\\
\hspace{-70pt}
~\quad
-2(-1)^{S-\Sigma'}
\sqrt{J(J+1)(2J+1)S(S+1)(2S+1)}
\nonumber\\
\hspace{-70pt}
\qquad\times
\sum_{q' =\pm 1}
\biggl[
\biggl(\begin{array}{ccc}
J & 2 & J \\
-\Omega' & -2q& \Omega'+2q
\end{array}\biggl)
\biggl(\begin{array}{ccc}
J & 1 & J \\
-\Omega'-2q & q' & \Omega
\end{array}\biggl)
\biggl(\begin{array}{ccc}
S & 1 & S \\
-\Sigma' & q' & \Sigma
\end{array}\biggl)
\theta(3/2-|\Omega'+2q|)
\nonumber\\
\hspace{-70pt}
\qquad\qquad\qquad
-
\biggl(\begin{array}{ccc}
J & 1 & J \\
-\Omega' & q' & \Omega'-q'
\end{array}\biggl)
\biggl(\begin{array}{ccc}
S & 1 & S \\
-\Sigma' & q' & \Sigma
\end{array}\biggl)
\biggl(\begin{array}{ccc}
J & 2 & J \\
-\Omega'+q' & -2q& \Omega
\end{array}\biggl)
\theta(3/2-| \Omega'-q'|)
\biggl]
\biggl\}
\, .\nonumber
\eeq
 
\textbullet~ $(P_D+2Q_D)$ term
\beq\fl
\lefteqn{
\bra{L\Lambda' \!,\! S\Sigma' \!,\! J'\Omega' M'_{J'} ,\! IM'_I}
\frac{P_D+2Q_D}{2}
\sum_{q=\pm 1}{\rm e}^{-2{\rm i}q\phi}
\left[
T^2_{2q}(\hat{\bf  J},\hat{\bf S})\hat{\bf N}^2+\hat{\bf N}^2T^2_{2q}(\hat{\bf  J},\hat{\bf S})
\right]
\ket{L\Lambda,\! S\Sigma, \! J\Omega M_J, \! IM_I}
}\nonumber\\
\hspace{-70pt}
=~
\frac{P_D+2Q_D}{2}
\delta_{J,J'}\delta_{M_J,M'_{J'}}\delta_{M_I,M'_{I}}
\sqrt{J(J+1)(2J+1)S(S+1)(2S+1)}
\sum_{q=\pm 1}
\delta_{\Lambda',\Lambda-2q}
\nonumber\\
\hspace{-70pt}
\times\biggl\{
(-1)^{J-\Omega'+S-\Sigma'}
\left[
2J(J+1)+2S(S+1)-2\Omega'\Sigma'-2\Omega\Sigma
\right]
\biggl(\begin{array}{ccc}
J & 1 & J \\
-\Omega' & -q& \Omega
\end{array}\biggl)
\biggl(\begin{array}{ccc}
S & 1 & S \\
-\Sigma' & q& \Sigma
\end{array}\biggl)
\nonumber\\
\hspace{-70pt}
~\quad
-2
\delta_{\Sigma,\Sigma'}
\sqrt{J(J+1)(2J+1)S(S+1)(2S+1)}
\theta(3/2-| \Omega'+q|)
\nonumber\\
\hspace{-70pt}
\qquad\times
\biggl(\begin{array}{ccc}
J & 1 & J \\
-\Omega' & -q & \Omega'+q
\end{array}\biggl)
\biggl(\begin{array}{ccc}
J & 1 & J \\
-\Omega'-q & -q& \Omega
\end{array}\biggl)
\nonumber\\
\hspace{-70pt}
\qquad\times
\biggl[
\biggl(\begin{array}{ccc}
S & 1 & S \\
-\Sigma & q& \Sigma-q
\end{array}\biggl)
\biggl(\begin{array}{ccc}
S & 1 & S \\
-\Sigma+q & -q & \Sigma
\end{array}\biggl)
+
\biggl(\begin{array}{ccc}
S & 1 & S \\
-\Sigma & -q & \Sigma+q
\end{array}\biggl)
\biggl(\begin{array}{ccc}
S & 1 & S \\
-\Sigma-q & q& \Sigma
\end{array}\biggl)
\biggl]
\biggl\}
\, .
\eeq

(14) Centrifugal distortion effect to $\Lambda$-doubling term (II)

\textbullet~ $-Q_H$ term
\beq\fl
\lefteqn{
-\bra{L\Lambda' \!,\! S\Sigma' \!,\! J'\Omega' M'_{J'} ,\! IM'_I}
\sum_{q=\pm 1}{\rm e}^{-2{\rm i}q\phi}\,
Q_H\frac{1}{2}
\left[
T^2_{2q}(\hat{\bf  J},\hat{\bf  J})(\hat{\bf N}^2)^2+(\hat{\bf N}^2)^2\,T^2_{2q}(\hat{\bf  J},\hat{\bf  J})
\right]
\ket{L\Lambda,\! S\Sigma, \! J\Omega M_J, \! IM_I}
}\nonumber\\
\hspace{-70pt}
=~
\frac{Q_H}{4\sqrt{6}}
\delta_{J,J'}\delta_{M_J,M'_{J'}}\delta_{M_I,M'_{I}}
\sqrt{(2J-1)2J(2J+1)(2J+2)(2J+3)}
\sum_{q=\pm 1}
\delta_{\Lambda',\Lambda-2q}
\nonumber\\
\hspace{-70pt}
\times
\biggl\{
\delta_{\Sigma,\Sigma'}
(-1)^{J-\Omega'}
\biggl[
\left[
J(J+1)+S(S+1)-2\Omega'\Sigma'
\right]^2
+
\left[
J(J+1)+S(S+1)-2\Omega\Sigma
\right]^2
\biggl]
\biggl(\begin{array}{ccc}
J & 2 & J \\
-\Omega' & -2q& \Omega
\end{array}\biggl)
\nonumber\\
\hspace{-70pt}
~\quad
+4
\delta_{\Sigma,\Sigma'}
(-1)^{J-\Omega'}
J(J+1)(2J+1)S(S+1)(2S+1)
\biggl(\begin{array}{ccc}
J & 2 & J \\
-\Omega' & -2q& \Omega
\end{array}\biggl)
\sum_{q'=\pm 1}
\biggl(\begin{array}{ccc}
S & 1 & S \\
-\Sigma & q'& \Sigma-q'
\end{array}\biggl)^2
\nonumber\\
\hspace{-70pt}
\qquad\times
\biggl[
\biggl(\begin{array}{ccc}
J & 1 & J \\
-\Omega' & q'& \Omega'-q'
\end{array}\biggl)^2
\theta(3/2-|\Omega'-q'|)
+
\biggl(\begin{array}{ccc}
J & 1 & J \\
-\Omega & q'& \Omega-q'
\end{array}\biggl)^2
\theta(3/2-|\Omega-q'|)
\biggl]
\nonumber\\
\hspace{-70pt}
~\quad
-2
(-1)^{S-\Sigma'}
\sqrt{J(J+1)(2J+1)S(S+1)(2S+1)}
\nonumber\\
\hspace{-70pt}
\qquad\times
\biggl[
\left[
2J(J+1)+2S(S+1)-2(\Omega'+2q)\Sigma'-2\Omega\Sigma
\right]
\biggl(\begin{array}{ccc}
J & 2 & J \\
-\Omega' & -2q& \Omega'+2q
\end{array}\biggl)
\theta(3/2-|\Omega'+2q|)
\nonumber\\
\hspace{-70pt}
\qquad\qquad\times
\sum_{q'=\pm 1}
\biggl(\begin{array}{ccc}
J & 1 & J \\
-\Omega'-2q & q'& \Omega
\end{array}\biggl)
\biggl(\begin{array}{ccc}
S & 1 & S \\
-\Sigma' & q'& \Sigma
\end{array}\biggl)
\nonumber\\
\hspace{-70pt}
\qquad\quad
-
\left[
2J(J+1)+2S(S+1)-2\Omega'\Sigma'-2(\Omega-2q)\Sigma
\right]
\biggl(\begin{array}{ccc}
J & 2 & J \\
-\Omega+2q & -2q& \Omega
\end{array}\biggl)
\theta(3/2-|\Omega-2q|)
\nonumber\\
\hspace{-70pt}
\qquad\qquad\times
\sum_{q'=\pm 1}
\biggl(\begin{array}{ccc}
J & 1 & J \\
-\Omega' & q'& \Omega-2q
\end{array}\biggl)
\biggl(\begin{array}{ccc}
S & 1 & S \\
-\Sigma' & q'& \Sigma
\end{array}\biggl)
\biggl]
\biggl\}
\, .
\eeq

\textbullet~ $(P_H+2Q_H)$ term
\beq\fl
\lefteqn{
\bra{L\Lambda' \!,\! S\Sigma' \!,\! J'\Omega' M'_{J'} ,\! IM'_I}
\sum_{q=\pm 1}\!{\rm e}^{-2{\rm i}q\phi}
\frac{P_H+2Q_H}{2}\!
\left[
T^2_{2q}(\hat{\bf  J},\hat{\bf S})(\hat{\bf N}^2)^2\!+\!(\hat{\bf N}^2)^2\,T^2_{2q}(\hat{\bf  J},\hat{\bf S})
\right]
\ket{L\Lambda,\! S\Sigma, \! J\Omega M_J, \! IM_I}
}\nonumber\\
\hspace{-70pt}
=~
\frac{P_H+2Q_H}{2}
\delta_{J,J'}\delta_{M_J,M'_{J'}}\delta_{M_I,M'_{I}}
\sqrt{J(J+1)(2J+1)S(S+1)(2S+1)}
\sum_{q=\pm 1}
\delta_{\Lambda',\Lambda-2q}
\nonumber\\
\hspace{-70pt}
\times
\biggl\{
(-1)^{J-\Omega'+S-\Sigma'}
\biggl(\begin{array}{ccc}
J & 1 & J \\
-\Omega' & -q& \Omega
\end{array}\biggl)
\biggl(\begin{array}{ccc}
S & 1 & S \\
-\Sigma' & q& \Sigma
\end{array}\biggl)
\nonumber\\
\hspace{-70pt}
~\qquad\times
\biggl[
\left[
J(J+1)+S(S+1)-2\Omega'\Sigma'
\right]^2
+
\left[
J(J+1)+S(S+1)-2\Omega\Sigma
\right]^2
\biggl]
\nonumber\\
\hspace{-70pt}
~\quad
+4
(-1)^{J-\Omega'+S-\Sigma'}
J(J+1)(2J+1)S(S+1)(2S+1)
\biggl(\begin{array}{ccc}
J & 1 & J \\
-\Omega' & -q& \Omega
\end{array}\biggl)
\biggl(\begin{array}{ccc}
S & 1 & S \\
-\Sigma' & q& \Sigma
\end{array}\biggl)
\nonumber\\
\hspace{-70pt}
~\qquad\times
\sum_{q'=\pm 1}
\biggl[
\biggl(\begin{array}{ccc}
J & 1 & J \\
-\Omega' & q' & \Omega'-q'
\end{array}\biggl)^2
\biggl(\begin{array}{ccc}
S & 1 & S \\
-\Sigma' & q' & \Sigma'-q'
\end{array}\biggl)^2
\theta(3/2-|\Omega'-q'|)
\nonumber\\
\hspace{-70pt}
~\qquad\qquad\qquad
+
\biggl(\begin{array}{ccc}
J & 1 & J \\
-\Omega & q' & \Omega-q'
\end{array}\biggl)^2
\biggl(\begin{array}{ccc}
S & 1 & S \\
-\Sigma & q' & \Sigma-q'
\end{array}\biggl)^2
\theta(3/2-|\Omega-q'|)
\biggl]
\nonumber\\
\hspace{-70pt}
~\quad
-2\delta_{\Sigma,\Sigma'}
\sqrt{J(J+1)(2J+1)S(S+1)(2S+1)}
\nonumber\\
\hspace{-70pt}
~\qquad\times
\biggl[
\left[
2J(J+1)+2S(S+1)-2(\Omega'+q)(\Sigma-q)-2\Omega\Sigma
\right]
\biggl(\begin{array}{ccc}
J & 1 & J \\
-\Omega' & -q& \Omega'+q
\end{array}\biggl)
\biggl(\begin{array}{ccc}
S & 1 & S \\
-\Sigma & q& \Sigma-q
\end{array}\biggl)^2
\nonumber\\
\hspace{-70pt}
~\qquad\qquad\times
\theta(3/2-| \Omega'+q|)
\biggl(\begin{array}{ccc}
J & 1 & J \\
-\Omega'-q & -q& \Omega
\end{array}\biggl)
\nonumber\\
\hspace{-70pt}
\qquad\qquad+
\left[
2J(J+1)+2S(S+1)-2\Omega'\Sigma-2(\Omega-q)(\Sigma+q)
\right]
\biggl(\begin{array}{ccc}
J & 1 & J \\
-\Omega+q & -q& \Omega
\end{array}\biggl)
\nonumber\\
\hspace{-70pt}
~\qquad\qquad\times
\theta(3/2-| \Omega-q|)
\biggl(\begin{array}{ccc}
J & 1 & J \\
-\Omega' & -q & \Omega-q
\end{array}\biggl)
\biggl(\begin{array}{ccc}
S & 1 & S \\
-\Sigma & -q & \Sigma+q
\end{array}\biggl)^2
\biggl]
\biggl\}
\, . 
\eeq

(15) Centrifugal distortion effect to magnetic hyperfine interaction
\beq\fl
\lefteqn{
\bra{L\Lambda' \!,\! S\Sigma' \!,\! J'\Omega' M'_{J'} ,\! IM'_I}
d_D\sum_{q=\pm 1} {\rm e}^{-2{\rm i}q\phi}\, 
\frac{1}{2}
\left[
T^2_{2q}(\hat{\bf I},\hat{\bf S})\hat{\bf N}^2+\hat{\bf N}^2T^2_{2q}(\hat{\bf I},\hat{\bf S})
\right]
\ket{L\Lambda,\! S\Sigma, \! J\Omega M_J, \! IM_I}
}\nonumber\\
\hspace{-70pt}
=~
\frac{d_D}{2}
\sqrt{(2J'+1)(2J+1)I(I+1)(2I+1)S(S+1)(2S+1)}
\nonumber\\
\hspace{-70pt}
\times
(-1)^{I-M'_{I} }
\sum_{p=0,\pm1}(-1)^{p}
\biggl(\begin{array}{ccc}
J' & 1 & J \\
 -M'_{J'} & p & M_J
\end{array}\biggl)
\biggl(\begin{array}{ccc}
I & 1 & I \\
 -M'_{I} & -p & M_I
\end{array}\biggl)
\sum_{q=\pm 1}
\delta_{\Lambda',\Lambda-2q}
\nonumber\\
\hspace{-70pt}
\times
\biggl\{
\left[
J'(J'+1)\!+\!J(J+1)\!+\!2S(S+1)\!-\!2\Omega'\Sigma'\!-\!2\Omega\Sigma
\right]
(-1)^{M'_{J'}-\Omega' +S-\Sigma'}
\biggl(\begin{array}{ccc}
J' & 1 & J \\
 -\Omega' & -q & \Omega
\end{array}\biggl)
\biggl(\begin{array}{ccc}
S & 1 & S \\
 -\Sigma' & q & \Sigma
\end{array}\biggl)
\nonumber\\
\hspace{-70pt}
~\quad
+2\delta_{\Sigma,\Sigma'}\theta(3/2-| \Omega'+q|)
(-1)^{J'+M'_{J'}}\sqrt{S(S+1)(2S+1)}
\nonumber\\
\hspace{-70pt}
~\qquad\times
\biggl[
\sqrt{J'(J'+1)(2J'+1)}
\biggl(\begin{array}{ccc}
J' & 1 & J' \\
-\Omega' & -q & \Omega'+q
\end{array}\biggl)
\biggl(\begin{array}{ccc}
S & 1 & S \\
-\Sigma & -q & \Sigma+q
\end{array}\biggl)
\nonumber\\
\hspace{-70pt}
~\qquad\qquad\times
\biggl(\begin{array}{ccc}
J' & 1 & J \\
 -\Omega'-q & -q & \Omega
\end{array}\biggl)
\biggl(\begin{array}{ccc}
S & 1 & S \\
 -\Sigma-q & q & \Sigma
\end{array}\biggl)
\nonumber\\
\hspace{-70pt}
\qquad\qquad
+(-1)^{J-J'}
\sqrt{J(J+1)(2J+1)}
\biggl(\begin{array}{ccc}
J' & 1 & J \\
 -\Omega' & -q & \Omega'+q
\end{array}\biggl)
\biggl(\begin{array}{ccc}
S & 1 & S \\
 -\Sigma & q & \Sigma-q
\end{array}\biggl)
\nonumber\\
\hspace{-70pt}
~\qquad\qquad\times
\biggl(\begin{array}{ccc}
J & 1 & J \\
-\Omega'-q & -q & \Omega
\end{array}\biggl)
\biggl(\begin{array}{ccc}
S & 1 & S \\
-\Sigma+q & -q & \Sigma
\end{array}\biggl)
\biggl]
\biggl\}
\, .
\eeq

(16) Orbital Zeeman effect 
\beq\fl
\lefteqn{
\bra{L\Lambda' \!,\! S\Sigma' \!,\! J'\Omega' M'_{J'} ,\! IM'_I}
g_L'\mu_B B_{Z} T^1_{p=0}(\hat{\bf L})
\ket{L\Lambda,\! S\Sigma, \! J\Omega M_J, \! IM_I}
}\nonumber\\
\hspace{-70pt}
=~
g_L'\mu_B B_{Z} 
\delta_{\Lambda,\Lambda'}
\delta_{\Sigma,\Sigma'}
\delta_{M_I,M'_{I}}
\Lambda
(-1)^{M'_{J'}-\Omega'} \sqrt{ (2J'+1)(2J+1)}
\biggl(\begin{array}{ccc}
J' & 1 & J \\
-\Omega' & 0 & \Omega 
\end{array}\biggl)
\biggl(\begin{array}{ccc}
J' & 1 & J \\
-M'_{J'} & 0 & M_J
\end{array}\biggl)
\, .
\eeq

(17) electronic spin isotropic contribution to Zeeman effect
\beq\fl
\lefteqn{
\bra{L\Lambda' \!,\! S\Sigma' \!,\! J'\Omega' M'_{J'} ,\! IM'_I}
g_S\mu_B B_{Z} T^1_{p=0}(\hat{\bf S})
\ket{L\Lambda,\! S\Sigma, \! J\Omega M_J, \! IM_I}
}\nonumber\\
\hspace{-70pt}
=~
g_S\mu_B B_{Z} 
\delta_{\Lambda,\Lambda'}\delta_{M_I,M'_{I}}
(-1)^{M'_{J'}-\Omega'+S-\Sigma'}
\sqrt{(2J'+1)(2J+1)S(S+1)(2S+1)}
\biggl(\begin{array}{ccc}
J' & 1 & J \\
-M'_{J'} & 0 & M_J
\end{array}\biggl)
\nonumber\\
\hspace{-70pt}
\quad\times 
\sum_{q=0,\pm1}
\biggl(\begin{array}{ccc}
J' & 1 & J \\
-\Omega' & q & \Omega
\end{array}\biggl)
\biggl(\begin{array}{ccc}
S & 1 & S \\
-\Sigma' & q & \Sigma
\end{array}\biggl)
\, .
\eeq

(18) Rotational magnetic moment contribution to Zeeman effect
\beq\fl
\lefteqn{
-
\bra{L\Lambda' \!,\! S\Sigma' \!,\! J'\Omega' M'_{J'} ,\! IM'_I}
g_r\mu_B B_{Z} T^1_{p=0}(\hat{\bf  J}-\hat{\bf L}-\hat{\bf S})
\ket{L\Lambda,\! S\Sigma, \! J\Omega M_J, \! IM_I}
}\nonumber\\
\hspace{-70pt}
=
-g_r\mu_B B_{Z}
\delta_{\Lambda,\Lambda'}\delta_{\Sigma,\Sigma'}
\delta_{J,J'}\delta_{\Omega,\Omega'}\delta_{M_J,M'_{J'}}
\delta_{M_I,M'_{I}}
 M_{J} 
\nonumber\\
\hspace{-70pt}~
+g_r\mu_B B_{Z}
\delta_{\Lambda,\Lambda'}
\delta_{\Sigma,\Sigma'}
\delta_{M_I,M'_{I}}
\Lambda
(-1)^{M'_{J'}-\Omega'} \sqrt{ (2J'+1)(2J+1)}
\biggl(\begin{array}{ccc}
J' & 1 & J \\
-\Omega' & 0 & \Omega 
\end{array}\biggl)
\biggl(\begin{array}{ccc}
J' & 1 & J \\
-M'_{J'} & 0 & M_J
\end{array}\biggl)
\nonumber\\
\hspace{-70pt}~
+g_r\mu_B B_{Z}
\delta_{\Lambda,\Lambda'}
\delta_{M_I,M'_{I}}
(-1)^{M'_{J'}-\Omega'+S-\Sigma'}
\sqrt{(2J'+1)(2J+1)S(S+1)(2S+1)}
\biggl(\begin{array}{ccc}
J' & 1 & J \\
-M'_{J'} & 0 & M_J
\end{array}\biggl)
\nonumber\\
\hspace{-70pt}
~\quad\times 
\sum_{q_1=0,\pm1}
\biggl(\begin{array}{ccc}
J' & 1 & J \\
-\Omega' & q_1 & \Omega
\end{array}\biggl)
\biggl(\begin{array}{ccc}
S & 1 & S \\
-\Sigma' & q_1 & \Sigma
\end{array}\biggl)
\, .
\eeq

(19) Nuclear spin Zeeman effect
\beq\fl
\lefteqn{
-
\bra{L\Lambda' \!,\! S\Sigma' \!,\! J'\Omega' M'_{J'} ,\! IM'_I}
g_N\mu_N  B_{Z} T^1_{p=0}(\hat{\bf I})
\ket{L\Lambda,\! S\Sigma, \! J\Omega M_J, \! IM_I}
}\nonumber\\
\hspace{-70pt}
=
-g_N\mu_N  B_{Z} 
\delta_{\Lambda,\Lambda'}\delta_{\Sigma,\Sigma'}
\delta_{J,J'}\delta_{\Omega,\Omega'}\delta_{M_J,M'_{J'}}
\delta_{M_I,M'_{I}}
M_{I}
\, ,\qquad\qquad
\qquad\qquad\qquad\qquad
\eeq
where the g-factor of Hydrogen nucleus is $g_N=2.792847$, 
the nuclear magneton is $\mu_N=e\hbar/m_p=5.05078343(43)\times10^{-27}\,{\rm J/T}$, 
and Bohr magneton is $\mu_B=e\hbar/m_e=9.27400915(23)\times10^{-24}\,{\rm J/T}$, 
which yields $\mu_N/\mu_B\simeq5.44617\times 10^{-4}$. 
We also remark that 
$1\mu_{B}\times 1\,{\rm Gauss}\simeq 1.39962\,{\rm MHz}$.

(20) electronic spin anisotropic contribution to Zeeman effect
\beq\fl
\lefteqn{
\bra{L\Lambda' \!,\! S\Sigma' \!,\! J'\Omega' M'_{J'} ,\! IM'_I}
g_{\ell}\mu_B B_{Z} \sum_{q=\pm1}\mathscr{D}_{0,q}^{(1)\ast}(\omega)T^1_{q}(\hat{\bf S})
\ket{L\Lambda,\! S\Sigma, \! J\Omega M_J, \! IM_I}
}\nonumber\\
\hspace{-70pt}
=~
g_{\ell}\mu_B B_{Z} 
\delta_{\Lambda,\Lambda'}
\delta_{M_I,M'_{I}}
(-1)^{M'_{J'}-\Omega'+S-\Sigma'}
\sqrt{(2J'+1)(2J+1)S(S+1)(2S+1)}
\biggl(\begin{array}{ccc}
J' & 1 & J \\
-M'_{J'} & 0 & M_J
\end{array}\biggl)
\nonumber\\
\hspace{-70pt}~\quad
\times 
\sum_{q=\pm1}
\biggl(\begin{array}{ccc}
J' & 1 & J \\
-\Omega' & q & \Omega
\end{array}\biggl)
\biggl(\begin{array}{ccc}
S & 1 & S \\
-\Sigma' & q & \Sigma
\end{array}\biggl)
\, .
\eeq

(21) Parity-dependent and non-cylindrical contribution to Zeeman effect (I)
\beq\fl
\lefteqn{
\bra{L\Lambda' \!,\! S\Sigma' \!,\! J'\Omega' M'_{J'} ,\! IM'_I}
g_{\ell}'\mu_B B_{Z} \sum_{q=\pm1}
{\rm e}^{-2{\rm i}q\phi}\,
\mathscr{D}_{0,-q}^{(1)\ast}(\omega)T^1_{q}(\hat{\bf S})
\ket{L\Lambda,\! S\Sigma, \! J\Omega M_J, \! IM_I}
}\nonumber\\
\hspace{-70pt}
=
-
g_{\ell}'\mu_B B_{Z}
\delta_{M_I,M'_{I}}
(-1)^{M'_{J'}-\Omega'+S-\Sigma'}
\sqrt{(2J'+1)(2J+1)S(S+1)(2S+1)}
\biggl(\begin{array}{ccc}
J' & 1 & J \\
-M'_{J'} & 0 & M_J
\end{array}\biggl)
\nonumber\\
\hspace{-70pt}~\quad
\times 
\sum_{q=\pm1}
\delta_{\Lambda',\Lambda-2q}
\biggl(\begin{array}{ccc}
J' & 1 & J \\
-\Omega' & -q & \Omega
\end{array}\biggl)
\biggl(\begin{array}{ccc}
S & 1 & S \\
-\Sigma' & q & \Sigma
\end{array}\biggl)
\, .
\eeq

(22) Parity-dependent and non-cylindrical contribution to Zeeman effect (II)
\beq\fl
\lefteqn{
-\!
\bra{L\Lambda' \!,\! S\Sigma' \!,\! J'\Omega' M'_{J'} ,\! IM'_I}
g_r^{e'}\mu_B B_{Z}\!\! \sum_{q=\pm1}\sum_{p=0,\pm1}\!\!\!
{\rm e}^{-2{\rm i}q\phi}
(-1)^p
\mathscr{D}_{-p,-q}^{(1)\ast}(\omega)T^1_{p}(\hat{\bf  J}\!-\!\hat{\bf S})
\mathscr{D}_{0,-q}^{(1)\ast}(\omega)
\ket{L\Lambda,\! S\Sigma, \! J\Omega M_J, \! IM_I}
}\nonumber\\
\hspace{-70pt}
=~
g_r^{e'}\mu_B B_{Z} 
\delta_{M_I,M'_{I}}
\sqrt{(2J'+1)(2J+1)}
\biggl(\begin{array}{ccc}
J' & 1& J \\
-M'_{J'} & 0 & M_J
\end{array}\biggl)
\sum_{q=\pm1}
\delta_{\Lambda',\Lambda-2q}
\nonumber\\
\hspace{-70pt}
\times\biggl\{
\delta_{\Sigma,\Sigma'}
(-1)^{J'+M'_{J'}}
\sqrt{J'(J'+1)(2J'+1)}
\biggl(\begin{array}{ccc}
J' & 1& J' \\
-\Omega' & -q & \Omega'+q
\end{array}\biggl)
\biggl(\begin{array}{ccc}
J' & 1& J \\
-\Omega'-q & -q & \Omega
\end{array}\biggl)
\theta(3/2-|\Omega'+q|)
\nonumber\\
\hspace{-70pt}
~\quad+
(-1)^{M'_{J'}-\Omega'+S-\Sigma'}
\sqrt{S(S+1)(2S+1)}
\biggl(\begin{array}{ccc}
J' & 1& J \\
-\Omega' & -q & \Omega
\end{array}\biggl)
\biggl(\begin{array}{ccc}
S & 1& S  \\
-\Sigma' & q & \Sigma
\end{array}\biggl)
\biggl\}
\, .
\eeq

(23) Stark effect
\beq\fl
\lefteqn{
\bra{L\Lambda' \!,\! S\Sigma' \!,\! J'\Omega' M'_{J'} ,\! IM'_I}
-\hat{\bf d}\cdot{\bf E}_{\rm DC}
\ket{L\Lambda,\! S\Sigma, \! J\Omega M_J, \! IM_I}
}\nonumber\\
\hspace{-70pt}
=
-\mu^{(e)}_z E_{\rm DC}
\delta_{\Lambda,\Lambda'}
\delta_{\Sigma,\Sigma'}
\delta_{M_I,M'_{I}}
(-1)^{M_{J'}'-\Omega'}
\sqrt{(2J'+1)(2J+1)}
\biggl(\begin{array}{ccc}
J' & 1 & J \\
-\Omega' & 0 & \Omega
\end{array}\biggl)
\nonumber\\
\hspace{-70pt}
~\quad\times
\sum_{p=0, \pm1}d^{(1)}_{p,0}(\theta_{BE})
\biggl(\begin{array}{ccc}
J' & 1 & J \\
-M_{J'}' & p & M_{J}
\end{array}\biggl)
\, .
\eeq

\section*{References}
\bibliography{KM_OH_ref.bib}

\end{document}